\documentclass[amsmath, amssymb, twocolumn, superscriptaddress, 10pt, aps, pra]{revtex4-1}

\usepackage{graphicx}
\usepackage{dcolumn}
\usepackage{bm}
\usepackage{xcolor}

\usepackage[utf8]{inputenc}
\usepackage[T1]{fontenc}
\usepackage{mathptmx}
\usepackage{etoolbox}
\usepackage{siunitx}

\usepackage{orcidlink}
\usepackage{url}

\begin{document}

\makeatletter
\def\@email#1#2{%
 \endgroup
 \patchcmd{\titleblock@produce}
  {\frontmatter@RRAPformat}
  {\frontmatter@RRAPformat{\produce@RRAP{*#1\href{mailto:#2}{#2}}}\frontmatter@RRAPformat}
  {}{}
}

\makeatother

\title{Extreme time extrapolation capabilities and thermodynamic consistency of physics-inspired Neural Networks for the 3D microstructure evolution of materials via Cahn-Hilliard flow}

\author{Daniele Lanzoni \orcidlink{0000-0002-1557-6411}$^{(*)}$} \email[Corresponding author: ]{daniele.lanzoni@unimib.it}
\affiliation{Materials Science Department, University of Milano-Bicocca, Via R. Cozzi 55, I-20125 Milano, Italy}

\author{Andrea Fantasia \orcidlink{0009-0009-1169-5083} }
\affiliation{Materials Science Department, University of Milano-Bicocca, Via R. Cozzi 55, I-20125 Milano, Italy}

\author{Roberto Bergamaschini \orcidlink{0000-0002-3686-2273} }
\affiliation{Materials Science Department, University of Milano-Bicocca, Via R. Cozzi 55, I-20125 Milano, Italy}

\author{Olivier Pierre-Louis \orcidlink{0000-0003-4855-4822}}
\affiliation{Institut Lumière Matière, UMR5306 Université Lyon 1—CNRS, 69622 Villeurbanne, France}

\author{Francesco Montalenti \orcidlink{0000-0001-7854-8269} }
\affiliation{Materials Science Department, University of Milano-Bicocca, Via R. Cozzi 55, I-20125 Milano, Italy}

\date{\today}

\begin{abstract}
A Convolutional Recurrent Neural Network (CRNN) is trained to reproduce the evolution of the spinodal decomposition process in three dimensions as described by the Cahn-Hilliard equation. A specialized, physics-inspired architecture is proven to provide close accordance between the predicted evolutions and the ground truth ones obtained via conventional integration schemes. The method can accurately reproduce the evolution of microstructures not represented in the training set at a fraction of the computational costs. Extremely long-time extrapolation capabilities are achieved, up to reaching the theoretically expected equilibrium state of the system, consisting of a layered, phase-separated morphology, despite the training set containing only relatively-short, initial phases of the evolution. Quantitative accordance with the decay rate of the Free energy is also demonstrated up to the late coarsening stages, proving that this class of Machine Learning approaches can become a new and powerful tool for the long timescale and high throughput simulation of materials, while retaining thermodynamic consistency and high-accuracy.
\end{abstract}

\maketitle

\section{Introduction}

Recently, there has been a surge in interest in Machine Learning (ML) methods~\cite{bishop2006pattern, goodfellowdeep2016} in the Computational Physics and Materials Science community~\cite{butler2018machine, Mehta2019PhysRep, bedolla2020machine, NGUYEN2024100544}. For instance, the new field of ML interatomic potentials~\cite{Bartok2010PRL, Kocer2022review} promises to push the boundaries of tractable time and spatial scales in molecular dynamics. An initially less explored route, which has however gained traction~\cite{Kim2019CGF, Fulton2019CGF, Zhang2020CMAME, MontesdeOcaZapiain2021npj, Yang2021Patterns, Lanzoni2022PRM, LUIS}, stems in the possibility of leveraging ML approaches, and Neural Networks (NN) in particular, for meso- and macroscopic models. Indeed, while it is easier to reach experimental scales using such tools, computational costs might still be critical for stiff problems and fine discretization requirements. ML may therefore provide an alternative to conventional solvers and allow one to bypass these issues. Indeed, in the last years, efforts have been dedicated to the possibility of approximating Partial Differential Equations (PDEs) with Machine Learning models~\cite{RAISSI2019PINN, Bretin2022JCompPhys}, with a particular interest on the morphological and microstructural evolution of materials~\cite{Yang2021Patterns, peivaste2022machine, Lanzoni2022PRM, fan2024accelerate}.

In this work, we explore the possibility of using a Convolutional Recurrent Neural Network (CRNN) to reproduce the time-dependent solutions of the Cahn-Hilliard equation, which describes the spinodal decomposition, an important process in binary mixtures leading to a spontaneous separation into two different phases~\cite{Langer1971AnnPhy, Kwon2007PRE, Andrews2020PRM, jinnai2000geometrical}, focusing on their long-time behavior. As the concentration fields can be conveniently mapped into a single order parameter, $\varphi$, the considered model also provides an initial test for the more general class of Phase Field (PF) models~\cite{Li2009CommunComput, provatas2011phase}. This is advantageous as PF models proved to be effective in several fields of Materials Science~\cite{chen2002phase, boettinger2002phase, Li2009CommunComput, provatas2011phase, albani}, thanks to their natural ability to tackle complex geometries, eventually involving topological changes, such as domain coalescence and splitting. Additionally, their formulation is open to the addition of multiple physical contributions, thus making the development of NN surrogates particularly appealing.

A 2D version of the spinodal decomposition has already been addressed via ML methods by the community in previous works, also using CRNNs~\cite{MontesdeOcaZapiain2021npj, Yang2021Patterns}. However, tackling the full 3D problem is essential to model realistic materials. For instance, the 3D coarsening of the spinodal decomposition may exhibit bi-continuous patterns, under suitable conditions, which cannot be observed in 2D. Additionally, full 3D simulations are computationally more challenging due to the requirement of a higher number of collocation points in discretization procedures, which further calls for acceleration with novel tools. Steps in this direction can be found in a few, very recent publications. In Ref.~\onlinecite{wang2023machine} a Convolutional (but not Recurrent) NN is shown to nicely predict relatively few steps of spinodal decomposition evolution, sufficient for the specific application therein. Ref.~\onlinecite{fan2024accelerate}, instead, shows that a Recurrent, Graph Network can surrogate grain coarsening dynamics in 2D and 3D. In the same work, an adaptive time-stepping scheme is proposed, which allows one to obtain an even larger speed-up in predictions. Still, only sequences of some hundred states are shown for the 3D case. In the present work, we therefore focus on the possibility of obtaining a model capable of producing stable, extremely long microstructural evolutions that are consistent with the underlying thermodynamic assumptions. This is done by exploiting a physics-inspired layer, which closely mimics the underlying material flow process in the ML model architecture. Importantly, predicted sequences are shown to reach the correct stationary state of the system, despite the CRNN only being presented orders of magnitude shorter, far from equilibrium sequences during training.

The paper is organized as follows. First, we briefly revise the spinodal decomposition model considered and we discuss the dataset creation procedure (Sec.~\ref{sec::pf&dataset}). Next, in Section~\ref{sec::NN&training} we define the NN method and we inspect training/validation performances. In Section~\ref{sec::simple_generalization}, we analyze the generalization capabilities of the CRNN, both in terms of computational cell size and length of the generated sequence. Lastly, a quantitative comparison between predictions and ground-truth evolution is performed, both in terms of local, voxel-level correspondence and global thermodynamic properties (Sec.~\ref{sec::quantitative}). Statistical significance is also tested.

\section{Methods}
\label{sec::methods}

\subsection{Phase Field model and dataset generation} \label{sec::pf&dataset}

Spinodal decomposition consists in the spontaneous separation of a binary mixture in two different phases below a critical temperature. From a mathematical standpoint, it is convenient to map the concentration fields tracking the local chemical composition to a single scalar field $\varphi$. This way, the dynamics of the system can be described in a diffuse interface, PF model~\cite{provatas2011phase}. The temporal evolution of the system may be derived from the definition of the free energy functional $F[\varphi]$:
\begin{equation}
    \label{eq::functional}
    F[\varphi] = \int_\Omega g(\varphi) + \frac{\varepsilon}{2} |\vec{\nabla} \varphi|^2 dx
\end{equation}
where $\Omega$ represents the physical domain considered, $g(\varphi)$ is the "bulk" contribution to the free energy, and the gradient term $|\vec{\nabla} \varphi|^2$ is related to interface energy and width $\varepsilon$~\cite{Li2009CommunComput, provatas2011phase}. As we are not interested in modeling a specific material, we choose for the bulk term $g$ the common and numerically convenient double-well potential expression $g(\varphi) = \frac{18}{\varepsilon} \varphi^2(1-\varphi)^2$~\cite{Li2009CommunComput}.

The equations of motion may be derived from a variational principle~\cite{provatas2011phase}. $\varphi$ flows in the direction of free-energy minimization, resulting in the well-known Cahn-Hilliard equation~\cite{cahn1958free, cahn1965JCP}:
\begin{equation}
    \label{eq::CH_equation}
    \frac{\partial \varphi}{\partial t} = - \vec{\nabla} \cdot \vec{J} =  M \nabla^2 (g'(\varphi) - \varepsilon \nabla^2 \varphi)
\end{equation}
where $g' = dg/d\varphi$, $M$ is a mobility constant and $\vec{J}$ is $\varphi$ diffusion flux, proportional to the gradient of the generalized chemical potential $\mu=\delta F/\delta \varphi$. Among PF models, the Cahn-Hilliard equation is particularly well suited as a relatively simple testing ground for advanced ML methods, as it can exhibit rich, non-linear behavior during evolution despite the relatively simple mathematical formulation, e.g. resulting in the formation and subsequent coarsening of 3D patterns, which are also interesting in the context of metamaterials design~\cite{kumar2020inverse, wang2023machine}.

The above-defined model is herein exploited to create a dataset of sequences of microstructure evolutions. Eq.~\ref{eq::CH_equation} is solved with a finite difference scheme on a cubic $64 \times 64 \times 64$ uniform grid using an explicit forward-Euler method with a constant integration step of $\delta t = 1.25 \times 10^{-3}$. $M$ is set to unity and $\varepsilon$ to 3 grid units. These parameters are chosen because they offer a good compromise between the accuracy of the solution of the PDE and the time required to generate the dataset. Periodic boundary conditions (PBCs) are imposed. As starting configurations for the field, we consider periodic Perlin noise samples~\cite{Perlin1985287}, obtained by suitable adaptation of the Python project at Ref.~\onlinecite{PythonPerlinNoise}. Perlin noise is a type of gradient noise that allows one to effectively select the typical correlation length of features in random $\varphi$ states. This is in contrast with the simpler initialization with uniform, random $\varphi$ values typically used, but provides coarser, grid-independent initial conditions that are easier for the CRNN to correlate with subsequent states of the evolution. Indeed, pure white noise cannot be recognized by CRNN as a sufficiently featured input, acting as an ill-defined initial condition and leading to arbitrary subsequent morphologies.

The generated training set comprises 1850 sequences with different average $\varphi$ values and correlation lengths, which are then split into a 70:30 ratio, forming the training and validation sets. The resulting average value of $\varphi$, $\overline{\varphi}$, has a bell-shaped distribution with mean $1/2$ and a standard deviation of $\approx 1/10$ as a result of the different Perlin noise parameters, ensuring the variability of initial conditions. Each sequence is composed of 50 consecutive configurations separated by a time interval $\tau = 0.5$ in time units (corresponding to 400 $\delta t$ integration steps). Training sequences start either from the initial Perlin noise or later states in the evolution, allowing the NN to generalize to partially coarsened microstructures as initial conditions.

\subsection{NN structure and training} \label{sec::NN&training}

\begin{figure*}[ht]
    \centering
    \includegraphics[width=\textwidth]{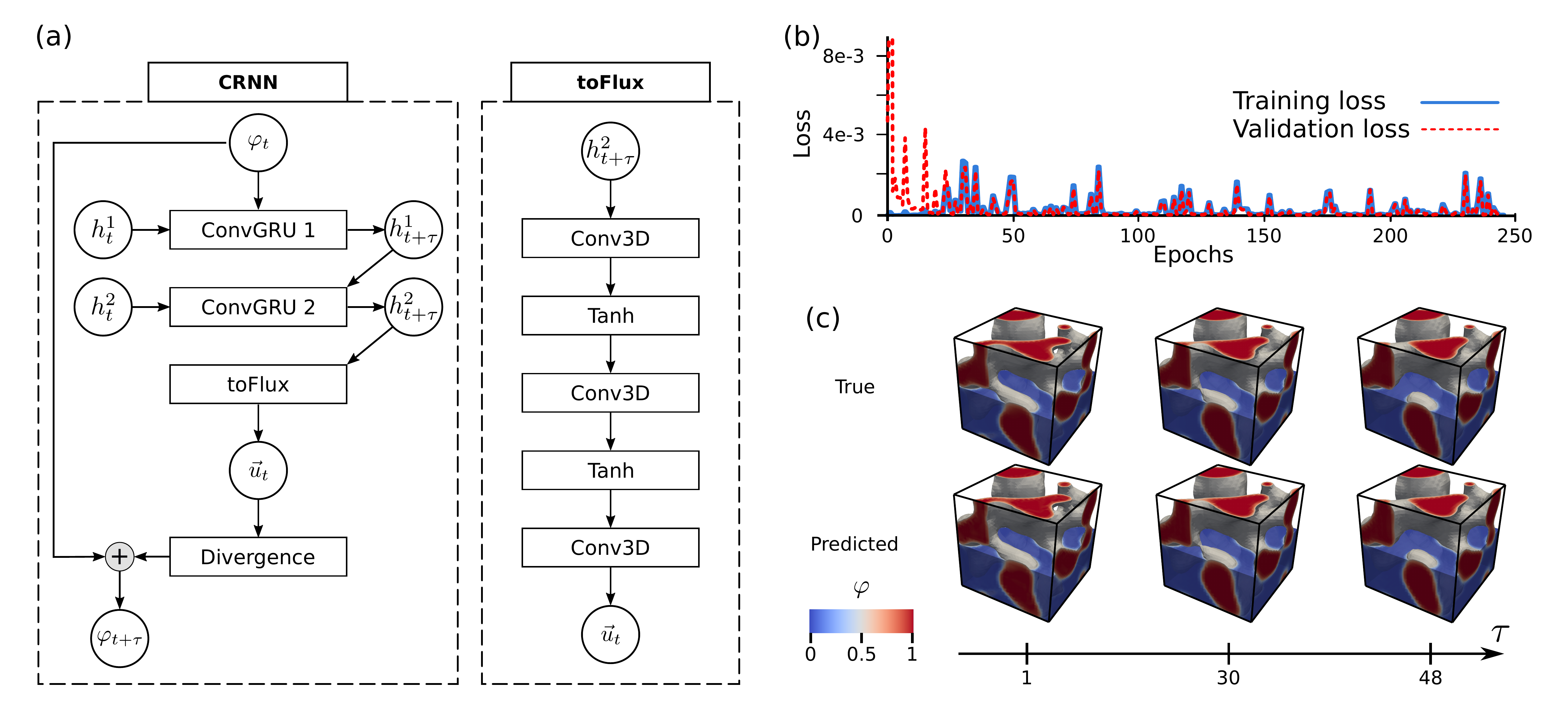}
    \caption{(a) Sketch of the NN architecture and specification of the "toFlux" layer converting the hidden state $h^2_{t+\tau}$ to the vector field $\vec{u}_t$. (b) Lossplot obtained training the Convolutional Recurrent Neural Network. No sign of overfitting is present. Differences in initial values for the curves are expected and due to the training procedure (see text for more information). (c) Comparison between the ground truth evolution and the predicted one for a validation set case. In the upper half of the snapshots, only $\varphi \ge 0.5$ is shown to better appreciate the internal structure.}
    \label{fig::lossplot}
\end{figure*}

The core architecture of the CRNN used in this work is based on Ref.~\onlinecite{Lanzoni2022PRM} but for three critical modifications. First, since we are considering 3D evolutions in the present work, all convolutional layers are promoted to their three-dimensional counterparts, using the corresponding PyTorch~\cite{paszkepytorch2019} implementation. Circular padding~\cite{schubert2019circular} enforces PBCs, but alternative boundary conditions could be straightforwardly translated into other padding modes. Second, a deeper mapping between the input, hidden and output layers in the recurrent modules of the CRNN (see GitHub repository for technical details) is exploited to increase the representation capabilities of the network. Third, a new output layer inspired by the physical formulation of the underlying problem is introduced. In particular, our main concern here is to strongly constrain the CRNN prediction to be consistent with conservative flow dynamics of Eq.\ref{eq::CH_equation}, a property that standard Convolutional layers do not possess. To this end, we use a simple forward-Euler integration scheme ansatz and, instead of directly predicting the next state of the field $\hat{\varphi}_{t+\tau}$ (hat denotes predicted quantities), we define the NN output as the vector field $\vec{u}_{t}$, such that:
\begin{equation}
    \label{eq::NN_integration}
    \hat{\varphi}_{t+\tau} = \hat{\varphi_t} + \vec{\nabla} \cdot \vec{u}_{t} .
\end{equation}
where $\vec{u}_t$ is a vector field acting as the flux term $-\tau \vec{J}_t$ in the standard forward-Euler integration scheme of Equation~\ref{eq::CH_equation}. Here $\vec{\nabla} \cdot \vec{u}_{t}$ is calculated using a finite difference scheme, efficiently implemented as non-trainable convolutional layers with circular padding and zero bias. Moreover, due to PBCs, $\vec{J}$ should have a vanishing mean, which in the NN architecture is translated as removing from $\vec{u}_{t}$ its mean value. While resembling a forward-Euler integration scheme, we remark that this approach is strictly more powerful, as the NN can exploit recurrence to access information farther in the evolution past, thus being more similar to a high-order integration scheme. More importantly, this procedure ensures the exact conservation of $\varphi$ by construction, thus making it suitable for other diffusive processes, such as the surface diffusion case of Ref.~\onlinecite{Lanzoni2022PRM}, where $\varphi$ conservation was enforced only weakly through the loss function. A similar strategy using an intermediate vector field was also exploited at Ref.~\onlinecite{Kim2019CGF} for incompressible flows.

This physics-inspired layer is in line with the well-known and used fact that good inductive biases encoded in the NN architecture lead to an easier training procedure, better generalization capabilities and a lower number of parameters, which in turn reduces overfitting risks~\cite{goodfellowdeep2016, Mehta2019PhysRep}. The present choice of fully convolutional layers~\cite{long2015fully} itself serves as an encoding of the translational equivariance symmetry~\cite{Cohen2016groupEquivariant} and of the locality of the evolution, allowing for applications to arbitrary domain sizes. In the Supplementary Material, it is shown how omitting the divergence layer degrades the predictive capabilities of the CRNN when extrapolating over long-time evolutions.

Based on our tests, the best set of hyperparameters is the following: the CRNN is composed of two stacked Convolutional Gated Recurrent Unit (GRU)~\cite{Chung20141arXiv, Shi2015NIPS} blocks, each using 10 channels for hidden states and $3 \times 3 \times 3$ convolution kernels. A sketch of the overall CRNN architecture is shown in Fig.~\ref{fig::lossplot}(a). The resulting NN model contains $\approx 2.9 \times 10^4$ trainable parameters, almost an order of magnitude less than the 2D prototype presented in Ref.~\onlinecite{Lanzoni2022PRM}, despite the more challenging 3D setting here tackled. Such compression is made possible by the modifications previously discussed. The training loss used is the voxel-wise Mean Squared Error between the predicted and ground truth sequence. Training of the NN parameters is performed using the standard implementation of the Adam optimizer~\cite{kingmaadam2014} with a learning rate of $10^{-5}$. Before being passed to the CRNN, training sequences are downscaled by a factor of $1/2$ in all spatial dimensions using nearest interpolation to ease GPU memory requirements during training. As it will be shown in the following, the NN prediction accuracy is not significantly impacted by such a downscale process, which, on the other hand, can be leveraged to tackle larger computational cells at reduced computational costs, as has also been demonstrated in previous works~\cite{Lanzoni2022PRM, fan2024accelerate}. Data augmentation based on reflections, \ang{90} degree rotations and the $\mathbf{Z}_2$ symmetry $\varphi \rightarrow (1-\varphi)$ is also exploited in training. We point again readers interested in implementation details to the GitHub repository, where the full code used to perform training and evolutions here reported is available.

Fig.~\ref{fig::lossplot}(b) reports training and validation loss respectively. To perform the full training, $\approx 50h$ are required using a workstation using a single Nvidia RTX A4000 GPU. Discrepancies between the two values in the initial training stages are due to the use of a curriculum learning schedule~\cite{bengio2009curriculum, goodfellowdeep2016}, which proved to speed up training convergence. Specifically, at the beginning of training the model parameters are optimized using a simplified loss function, i.e. the CRNN is required to generate only the last state provided the previous 49 ones. In the following 25 epochs, the loss function gets gradually more complex as the CRNN is required to generate progressively longer sequences. At the end of this curriculum stage, the ML model has to reproduce the full sequence based only on the initial condition.

\begin{figure}[ht]
    \centering
    \includegraphics[width=\columnwidth]{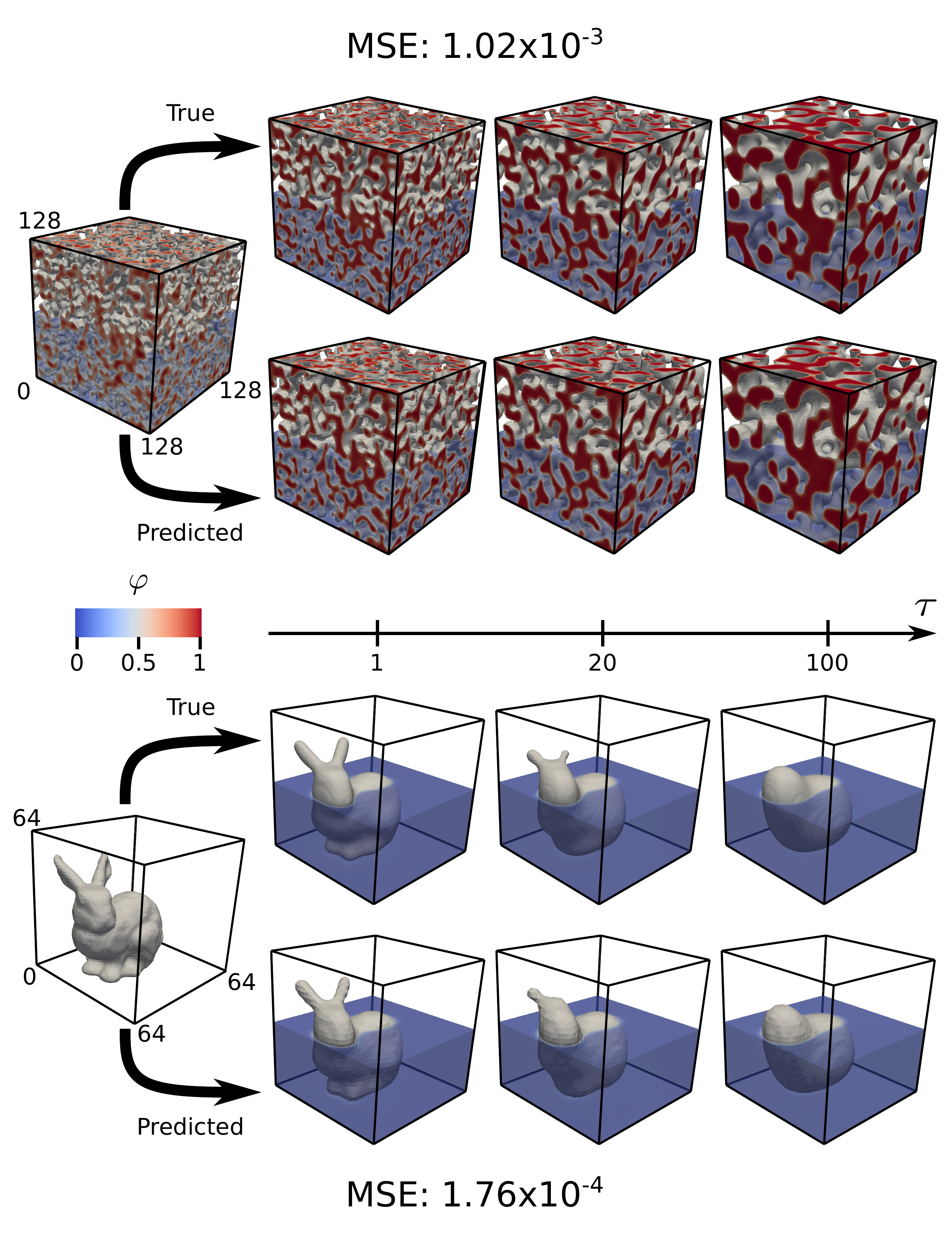}
    \caption{(a) Spatial generalization test. The domain size is 8 times larger than that used in the training set (twice as large in all directions). One-to-one accordance with the ground truth evolution can still be observed. (b) Evolution on a computational cell with the same size of training for a $\varphi$ initial profile shaped as the Stanford bunny. In the upper half of the snapshots, only $\varphi \ge 0.5$ is shown. MSE values for the whole sequences are reported in the insets.}
    \label{fig::spatial_extra}
\end{figure}

The validation loss does not increase in the late stages of training, indicating that no overfitting is present. As both training and validation losses present spikes, the model with the lowest validation loss in the last 50 epochs has been selected. A comparison between the ground-truth finite difference evolution and the one provided by the NN for a validation case is reported in Fig.~\ref{fig::lossplot}(c). One-to-one correspondence in the $\varphi$ maps can be observed. Notice that the typical variation between subsequent $\tau$ intervals in the training evolution is fairly localized, which further justifies the Euler-like ansatz of Equation~\ref{eq::NN_integration}.

\section{Results}
\label{sec::results}

As a good practice, after the training and validation procedure, we check the model predictivity on an independent test set with similar characteristics to the examples used during the optimization. The obtained results present the almost one-to-one correspondence already observed for the validation set in Fig.~\ref{fig::lossplot}(c) and are therefore not reported. Instead, we move directly to more challenging cases, showing how the CRNN can generalize to larger computational domains and generate far longer sequences than those observed in training. In the following, we will inspect generated sequences (Sects.~\ref{sec::simple_generalization} and~\ref{sec::stationary}), discussing the quantitative aspects in the last Section (\ref{sec::quantitative}).

\subsection{Domain size and evolution time generalization tests} \label{sec::simple_generalization}

One of the main advantages of a fully convolutional NN is the possibility of applying the trained model to inputs of arbitrary size. This is made possible thanks to the implicit assumption of locality embedded in the Convolutional layer choice. While for non-local PDEs this kind of procedure should be considered with care, the current application to the Cahn-Hilliard equation as defined in Eq.~\ref{eq::CH_equation} does not pose any issues. To check the generalization capabilities of our CRNN in this setting, we report in Fig.~\ref{fig::spatial_extra}(a) a computational domain twice as large in all spatial dimensions and evolved for double the amount of time steps used for training examples. It can be observed that the NN prediction closely reproduces the finite difference evolution at all reported stages. The MSE loss on the whole sequence is also reported, exhibiting a value slightly higher than training and validation.

\begin{figure}[ht]
    \centering
    \includegraphics[width=\columnwidth]{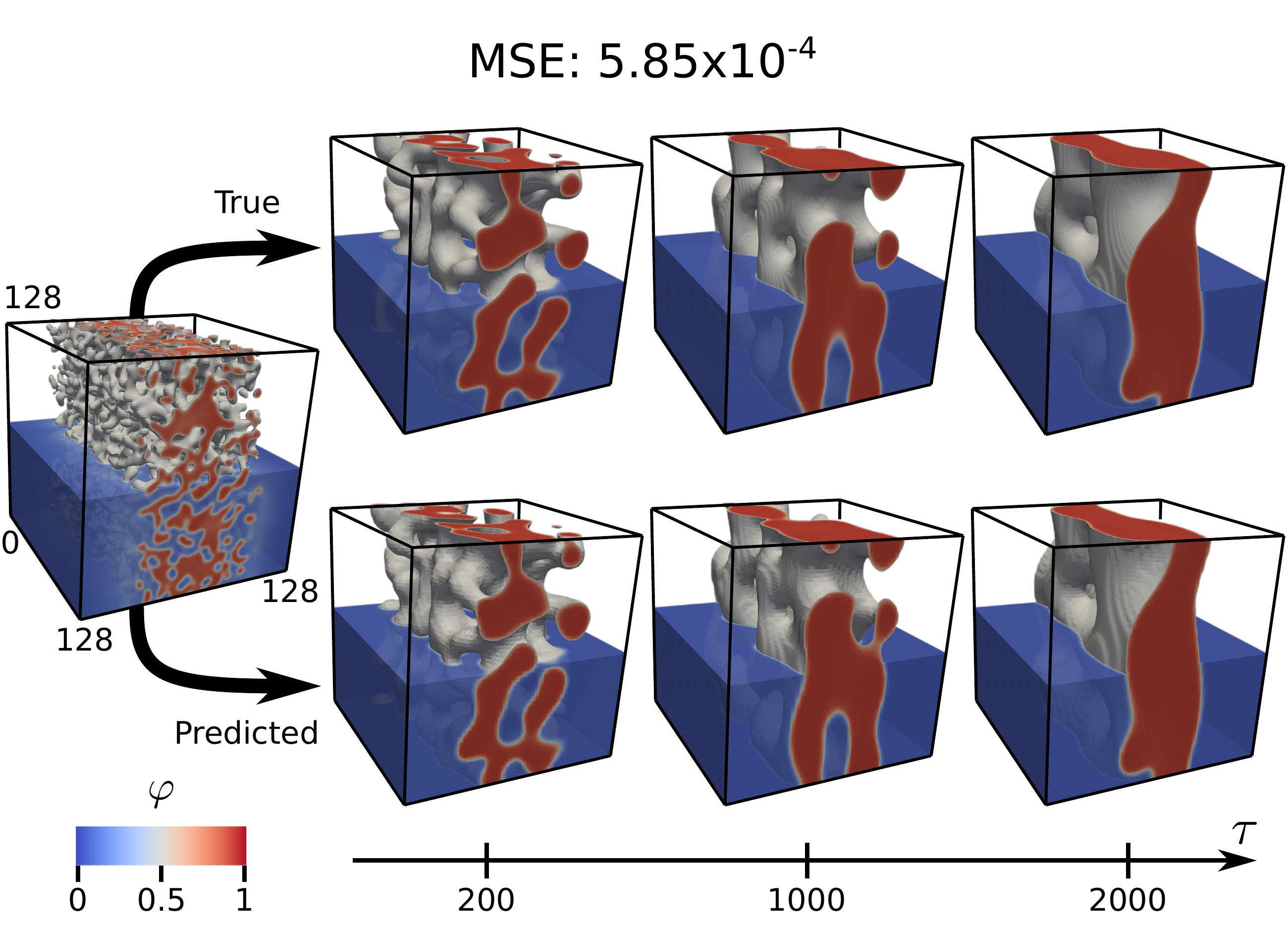}
    \caption{Time generalization test. A domain enclosing a region in which $\varphi$ rapidly fluctuates is evolved for $40$ times more steps than those used in training. Despite local variations, long-time behavior still exhibits quantitative correspondence. In the upper half of the snapshots, only $\varphi \ge 0.5$ is shown. MSE loss for the whole sequence is reported in the inset.}
    \label{fig::time-extra}
\end{figure}

As an additional proof of the CRNN generalization capabilities on qualitatively different configurations, in Fig.~\ref{fig::spatial_extra}(b) we also report the evolution under the Cahn-Hilliard flow of a domain initialized with the shape of the Stanford bunny~\cite{StanfordBunny}, which is clearly out of the distribution of typical training set microstructures. The domain size, in this case, is the same as the $64 \times 64 \times 64$ grid used in training, but coarsening stages at 100 $\tau$ are inspected as in panel (a). Close correspondence with the ground truth evolution is again achieved, despite some local variations (e.g. the discrepancy at 20 $\tau$ in the left ear), which however are not preventing the close correspondence at later stages. The MSE loss, reported at the bottom of Fig.~\ref{fig::spatial_extra} is comparable to values in training/validation, confirming the excellent generalization capabilities of the CRNN.

We remark that, similarly to a simple forward-Euler integration scheme, computational costs scale linearly with the number of collocation points, albeit with a longer time interval between subsequent predicted states. We point the technical reader to Appendix~\ref{sec::computational_costs} where this aspect is quantitatively analyzed.

\begin{figure*}[ht]
    \centering
    \includegraphics[width=\textwidth]{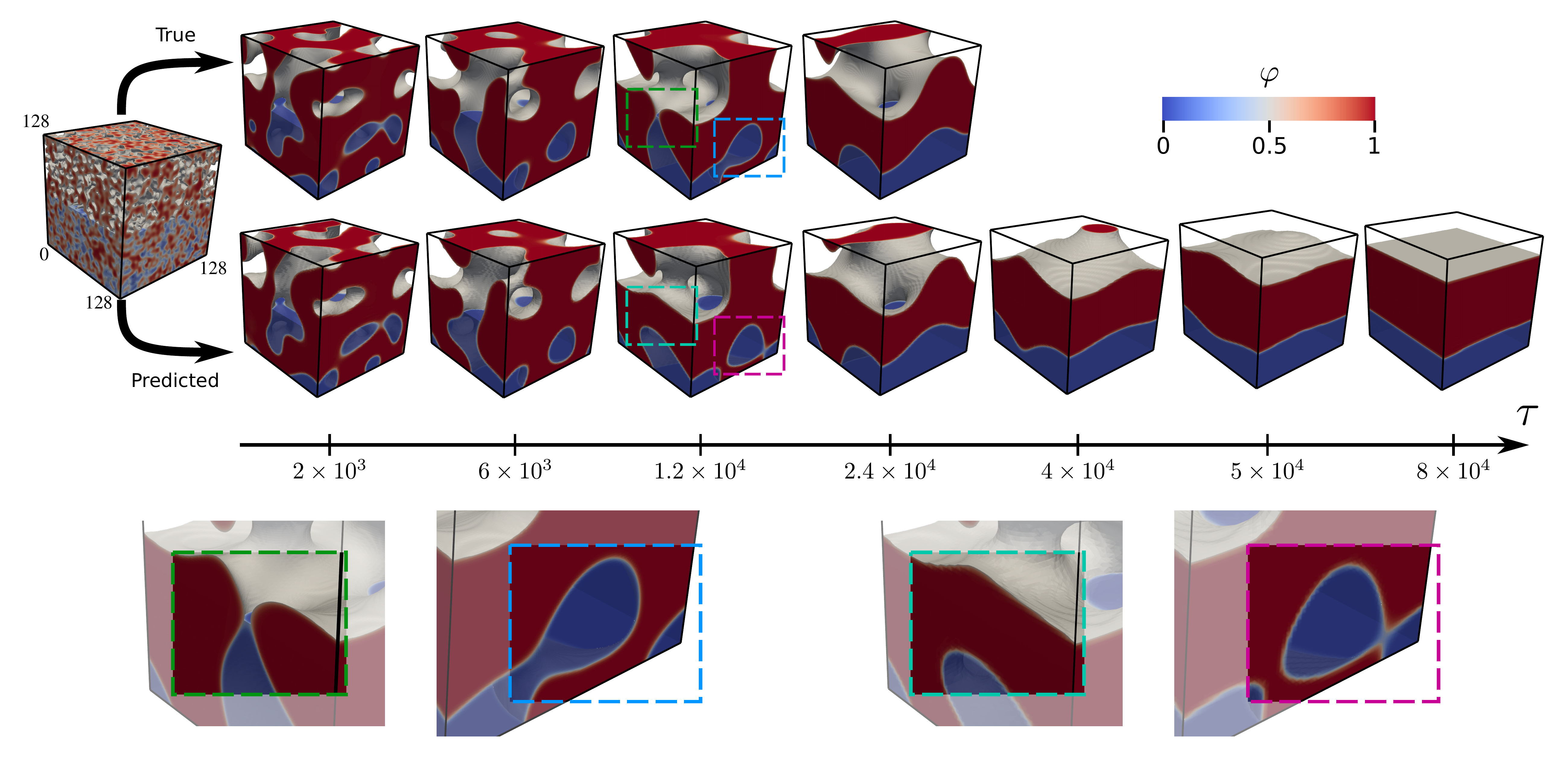}
    \caption{Extreme time generalization test. A domain initialized with Perlin random noise is evolved until a flat configuration is reached, consistently with one of the possible stationary states for the system. A comparison with the evolution obtained using the finite difference integration is provided for $\approx 1/3$ of the sequence. In the upper half of the snapshots, only $\varphi \ge 0.5$ is shown. Close-ups of regions presenting the largest local deviations at time $1.2 \times 10^4 \tau$ are also shown in the insets.}
    \label{fig::stationary}
\end{figure*}

Another generalization test regards the possibility of generating sequences of arbitrary length by iteratively re-feeding the CRNN's own output as input to generate the following evolution stage. This aspect is particularly critical, as this recurrence may lead to prediction error accumulation and loss of reliability in the generated sequence. In principle, therefore, one could expect the evolution obtained by NN and the one coming from finite difference integration to diverge gradually. Furthermore, as dynamics progress, predicted morphologies arising from domain coarsening might be less represented in the training set. This could lead to particularly severe limitations for ML approaches, as configurations emerging from the considered Phase Field model are subject to several constraints, in particular the conservation and boundedness of $\varphi$. Notice that, while local conservation is ensured (up to numerical precision) by the specialized, physics-inspired layer ("toFlux" and divergence operation in Fig.~\ref{fig::lossplot}(b)), there is no guarantee, so far, that very long sequences will not produce invalid phase field representations, e.g. with diverging $\varphi$ values. To demonstrate the generalization capabilities of the proposed CRNN, in Fig.~\ref{fig::time-extra}(a) we show the comparison between the predicted and ground truth evolution for an initial configuration with fluctuations localized in the central slice of a $128 \times 128 \times 128$ grid. It can be seen that there is a substantial agreement between such "long-time" predictions from the NN and the true evolutions provided by the explicit integration scheme. Once again, despite local differences, the overall sequence and the final state of the evolution exhibit compelling correspondence, with an MSE loss of the same order of magnitude as the one observed in the top evolution of Fig.~\ref{fig::spatial_extra}. Notice that the complete evolution requires $8 \times 10^{5}$ finite difference integration steps, which correspond to a sequence $40$ times longer than those contained in the training set.

\subsection{Stationary state prediction}
\label{sec::stationary}

As stated in previous Sections, CRNN approaches are particularly convenient from a computational standpoint, as numerical operations can be easily parallelized on GPUs and multi-threaded systems. This allows for a strong compression of simulation wall times. Regarding examples in previous sections, the speedup obtained on an Nvidia RTX A4000 GPU running on the same machine that performed the (non-optimized) Phase Field simulations was of $\approx 10^3$. This is only an indicative figure of merit, as, for instance, more advanced integration schemes could significantly speed up spinodal decomposition simulations~\cite{CHEN1998PFpseudospectral}. On the other hand, a similar NN approach could be applied to more complex PDEs, where such advanced schemes are possibly not available or not as advantageous.

We now take profit from this computational speedup to inspect what happens if the NN model is run to generate sequences orders of magnitude longer than those provided in training. In particular, our scope is here to investigate whether the error accumulation eventually leads the NN predictions to un-physical configurations and whether the predicted stationary states (if any) are consistent with those expected from the Cahn-Hilliard model. Fig.\ref{fig::stationary} reports the NN-predicted sequence of microstructure coarsening stages going from a Perlin-noise profile (with similar parameters to those used in training but not present in the dataset) to its stationary state, together with a one-to-one comparison with the corresponding finite difference integration up to $\approx 1/3$ of the evolution. Notably, the generated sequence requires $8 \times 10^4$ NN iterations, which is 4 orders of magnitude longer than those provided in training. A full simulation using the explicit integration scheme therefore requires more computational resources than those used for the whole dataset building.

Regarding error accumulation, local discrepancies reduce as dynamics progress, as expected from the Cahn-Hilliard flow minimizing the interface area between the two phases. Indeed, small dimples or corrugations in a lower curvature interface tend to be flattened by the evolution law defined by Eq.~\ref{eq::CH_equation}, as can be observed in the initial stages of the evolution (see e.g. fine details in the Stanford bunny in Fig.~\ref{fig::spatial_extra}). Inspecting the configurations at time $1.2 \times 10^4 \tau$ of Fig.~\ref{fig::stationary}, it is possible to notice that the connectivity of the red domain shows significant deviations (see also close-ups provided in the insets at the bottom of the figure). At the time $2.4\times 10^4 \tau$, however, the difference between the two configurations is far less severe, showing that the NN prediction has closed the gap with the ground-truth evolution. This testifies that the CRNN has correctly learned the underlying PF model and that no net error accumulation is pushing the predictions systematically toward un-physical configurations. This can be, at least in part, ascribed to the Recurrent nature of the NN: since the model is trained to maximize the likelihood of a whole sequence, temporary discrepancies are allowed, as long as the overall prediction gets closer again to the ground truth in the long run. A quantification of this effect is provided in Sect.~\ref{sec::quantitative}.

\begin{figure}
    \centering
    \includegraphics[width=\columnwidth]{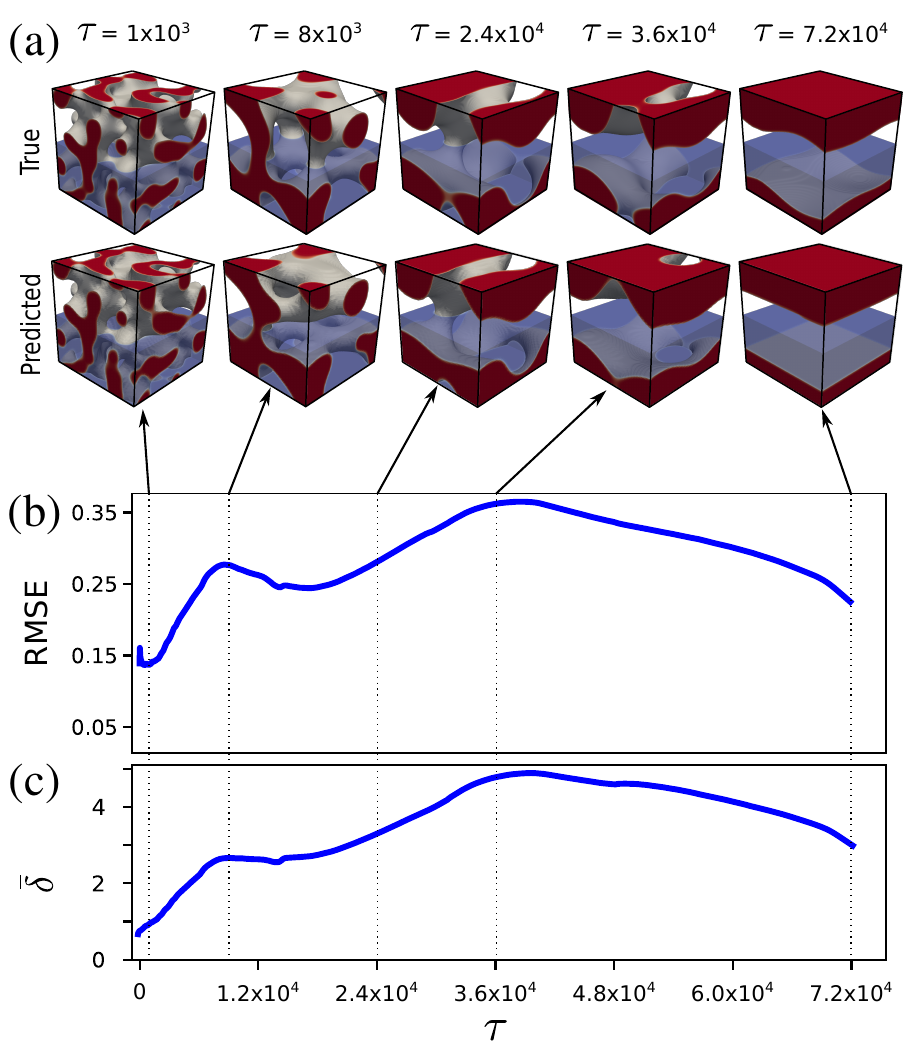}
    \caption{(a) Representative stages comparing the ground-truth evolution and the corresponding NN prediction. (b) RMSE curve between the predicted and ground truth $\varphi$. (c) $\bar{\delta}$ error (Eq.~\ref{eq::measure}) as a function of time.}
    \label{fig::phi_quantification}
\end{figure}

Fig.~\ref{fig::stationary} also reports a stationary state presenting a layered microstructure, which is reached at time $\approx 8 \times 10^4 \tau$ and corresponds to a time sequence $1600$ times longer than those observed in training. Importantly, such configuration is stable once reached, despite being out of the distribution of the examples in the dataset. This is a striking proof of the generalization capabilities of the CRNN, as this layered microstructure is one of the expected stationary solutions of the Cahn-Hilliard equation under the Periodic Boundary Conditions considered here~\cite{provatas2011phase}.

\subsection{Quantitative evaluation of predictive performances}
\label{sec::quantitative}

Until this point, the assessment of the NN performances was only qualitative and mainly based on visual inspection. This last section addresses this issue by performing a more quantitative analysis. The long-time behavior of the system is investigated using a spinodal decomposition simulation on a $128 \times 128 \times 128$ grid, starting from a Perlin noise initial condition and $7.2 \times 10^4 \tau$ long, compatible with the one already shown in Fig.~\ref{fig::stationary}. The comparison between representative stages of the two sequences is reported in Fig.~\ref{fig::phi_quantification}(a). To quantify the difference between the two evolutions, we report the Root Mean Squared Error (RMSE), here evaluated as ($V$ is the computational domain volume)
$$
\text{RMSE}(\varphi, \hat{\varphi}) = \sqrt{ 1/V \int_\Omega (\varphi - \hat{\varphi})^2 dx } \ 
$$
and reported in Fig.~\ref{fig::phi_quantification}(b).

Notice that, despite the initial condition being the same for both the NN and the ground truth evolution, the RMSE starts from a non-vanishing value. The origin of such behavior is due to the artifacts introduced by the downscaling interpolation required by the NN. The initial RMSE of $\approx 0.13$ should therefore be considered "intrinsic" to the downscaling procedure, but it does not affect the subsequent evolution. In Appendix~\ref{sec::downscaling_effect}, we explicitly verify this by calculating the same quantity after interpolation on a $64 \times 64 \times 64$ grid. In later evolution stages, the RMSE increases and decreases again, e.g. at $\tau \approx 1.8 \times 10^4$ and at the end of the reported evolution, when both the NN prediction and the ground truth evolution reach a layered microstructure. This shows that the accumulation of deviations is progressively recovered, as expected from the Cahn-Hilliard evolution. It is also interesting to note that the time required to reach a two-interface microstructure is consistent between the true and predicted evolution, although the CRNN is faster in smoothing interface undulations.

While the RMSE is closely related to the loss function used in training, it lacks an intuitive meaning of its numerical value. To quantify the prediction error with a variable with more direct physical interpretation, we also report in Fig.~\ref{fig::phi_quantification} the quantity $\bar{\delta}$, defined as:
\begin{equation}
    \bar{\delta} = \sqrt{ \frac{\int_\Omega (\varphi - \hat{\varphi})^2 dx}{\int_\Omega |\vec{\nabla} \varphi|^2 dx} }.
    \label{eq::measure}
\end{equation}
If the ground truth $\varphi$ and the prediction $\hat{\varphi}$ differ by a small displacement of the position of the interface between the two phases, Eq.~\ref{eq::measure} yields the root mean square value of such displacement. Appendix~\ref{sec::measure} reports a more complete discussion and a mathematical derivation. $\bar{\delta}$ can therefore be interpreted as the average, local distance between the predicted and true interface in grid units. Similarly to what was observed for the RMSE, this value increases in the intermediate stages and decreases towards the end of the reported dynamics, as the microstructure coarsens forming interfaces parallel to the computational domain sides. The initial condition high value in the RMSE is far less pronounced in terms of $\bar{\delta}$, but notice that this value refers to a random noise configuration lacking well-defined interfaces. We recall that the interface thickness parameter $\varepsilon = 3$ (see Eq.~\ref{eq::CH_equation}), which means that the root mean square distance between the interfaces is comparable for most of the evolution to the width of transition regions between the two phases.

\begin{figure*}
    \centering
    \includegraphics[width=\textwidth]{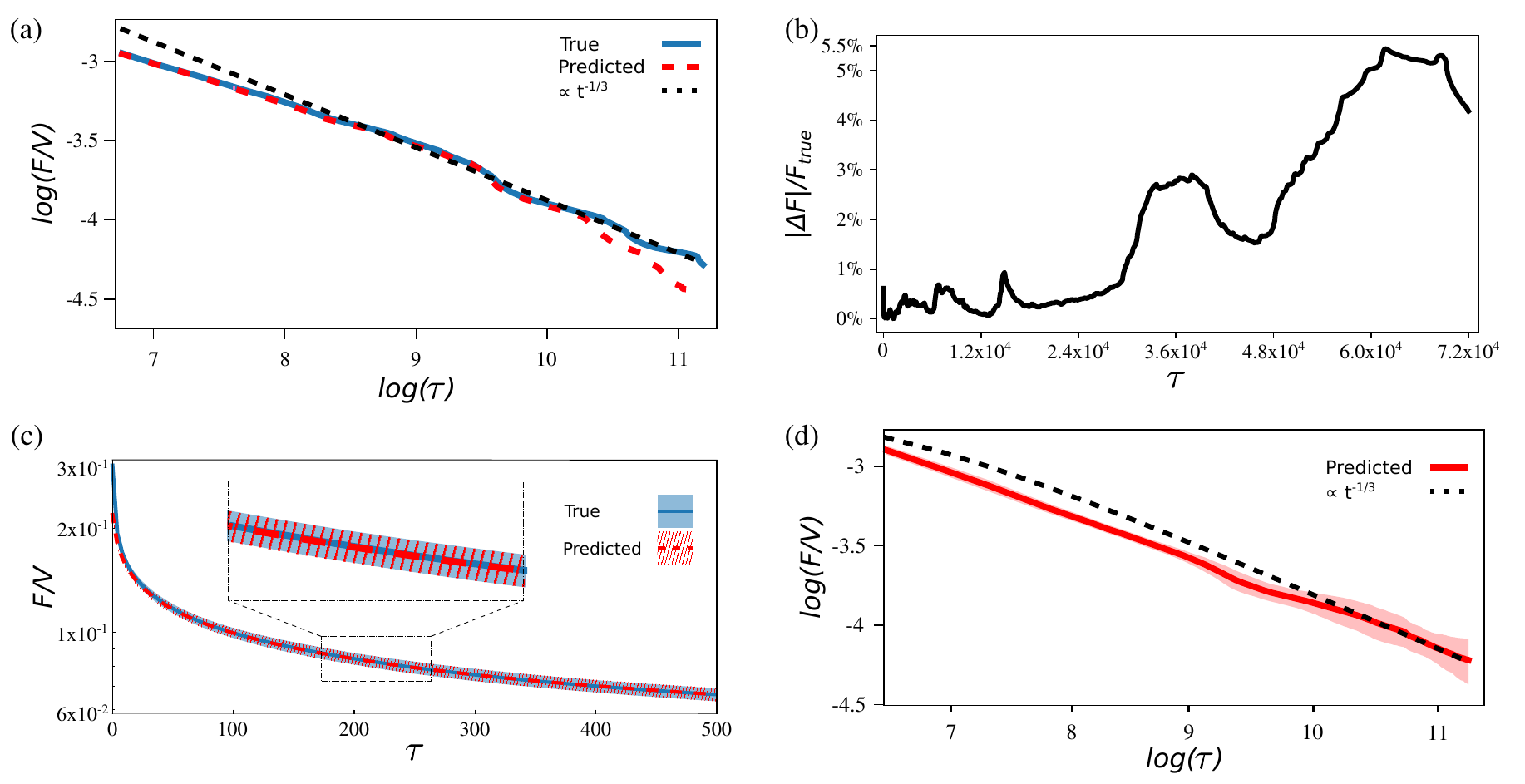}
    \caption{(a) Predicted and ground-truth free energies per unit volume $F/V$ for the same time sequences of Fig.~\ref{fig::phi_quantification}. (b) Relative free energy difference between the two evolutions (linear scale). (c) Free energy density $F/V$ decay curves obtained by averaging out an ensemble of 100 independent evolutions initialized by Perlin noise and obtained by explicit integration (blue solid curve) and CRNN (red dashed curve). Shaded areas report the corresponding single-standard deviation range. (d) Free energy density $F/V$ decay curved obtained by averaging 10 independent evolutions by CRNN. The shaded area corresponds to one standard deviation.}
    \label{fig::simple_quantification}
\end{figure*}

A point-by-point comparison of the evolution yields quantitative information on the NN prediction on a fine level. Indeed, both the RMSE and $\bar{\delta}$ are sensitive to local variations. However, they do not provide information on how much the predicted sequences are globally consistent with the underlying physical mechanisms. In this respect, it is crucial to assess whether the CRNN prediction remains consistent with the Cahn-Hilliard flow even when the RMSE or $\bar{\delta}$ values are at their maxima. Fig.~\ref{fig::simple_quantification}(a) reports the time dependence of the free energy of the system normalized on the domain volume $F/N$ for both the NN prediction and the corresponding ground-truth evolution. A logarithmic scale is used to appreciate finer details and to analyze the power-law trends of free-energy decay. In panel (b), the relative absolute error is also reported (linear scale is used in this case). It is evident how the predicted evolution of the free energy closely reproduces the "true" curve, even when the error values of Fig.~\ref{fig::phi_quantification}(a) increase, testifying that the CRNN predictions follow the underlying free-energy minimization process. The largest deviations are observed near the late stages of the evolution ($log(\tau) \ge 10.25$, i.e. $\tau \ge 2.8\times 10^4$), which exhibit an acceleration in the coarsening rate with respect to the true evolution. Nonetheless, the relative absolute error in the predicted $F$ stays below 5.5\% for the whole evolution, as shown in Fig.~\ref{fig::phi_quantification}(b).

Fig.~\ref{fig::simple_quantification}(a) also allows for a second quantitative analysis, regarding the coarsening rate of the microstructure. In Cahn-Hilliard evolution the typical domain length is expected to grow in time as $t^{1/3}$ and, as a consequence, the interface area per unit volume is expected to decay as $t^{-1/3}$~\cite{kohn2002upper, kwon2010morphology}. From the definition of $F$ in Eq.~\ref{eq::functional}, in this work, the free energy corresponds to the interface area between the two phases. We therefore expect that $F(t) \propto t^{-1/3}$ in the long-time regime. The dashed black line reported in Fig.~\ref{fig::simple_quantification}(a) corresponds to such behavior. Clearly, the ground truth $F(t)$ line is parallel to the expected one, indicating the same dependence for the free energy decay. Local deviations and fluctuations are related to finite domain size effects and interface splitting/merging processes. The NN prediction also follows the same power-law, up to $log(\tau) \approx 10.25$ (again, corresponding to $\tau \approx 2.8 \times 10^4$), where the coarsening rate increases with respect to the ground truth and gets closer to a $\propto t^{-0.44}$ dependence (not shown). This behavior is probably due to the lack of long-time configurations in the training set and may be eliminated with active or transfer learning workflows, the implementation of which we leave for future works. Remarkably, while not completely satisfactory, this does not prevent the NN from predicting the correct stationary state.

To confirm the statistical significance of the above discussion, we perform a similar analysis on an ensemble of 100 evolutions, obtained using a $64 \times 64 \times 64$ grid and Perlin-noise independent initial configurations consistent with those in the training set. A total time of $\tau=500$ is considered, as the computational effort required makes it unpractical to consider longer sequences. Fig.~\ref{fig::simple_quantification}(c) reports in semi-logarithmic scale the decay of the free energy obtained by averaging across all evolutions, along with its variability range, represented by the standard deviation and shown as the shaded area. A close-up of the two curves is also shown to better appreciate fine details. As it can be seen, predicted and ground-truth quantities are in close agreement, both in terms of means and standard deviations, with the largest difference in the initial stages, where the effects of the random initialization are stronger. These results confirm that the generated microstructures are quantitatively consistent with the underlying Cahn-Hilliard material flow also on an ensemble level.

We close the discussion by addressing the problem of the wrong exponent in the free energy dissipation observed at the end of the evolution in Fig.~\ref{fig::simple_quantification}(a). While performing an extensive CRNN-ground truth comparison is computationally demanding, we can attack the problem indirectly by exploiting the low cost of Machine Learning predictions. Fig.~\ref{fig::simple_quantification}(d) reports in logarithmic scale the average free energy decay curve over 10 independent evolutions $80000 \tau$ long on a $128 \times 128 \times 128$ grid, each starting from a different Perlin noise initial condition. One standard deviation interval is also reported with the red-shaded area. As it can be clearly appreciated, the average long-time behavior exhibits the expected $t^{-1/3}$ power law, showing that the acceleration previously reported is not systematic. It may therefore be concluded that the NN predictions do not exhibit a net, strong deviation of the expected behavior of $F(t)$, despite the relatively short sequences used in training.

\section{Conclusions}

This work shows that Convolutional Recurrent Neural Networks provide an excellent approach to approximate the microstructural evolution of materials undergoing the spinodal decomposition process in three dimensions. The use of fully convolutional and recurrent structures, together with a physics-inspired specialized layer, yields the accurate generation of spatiotemporal sequences at a fraction of the computational costs of the explicit method used to build the training set. Close correspondence for evolutions much longer than those provided in training has been demonstrated. Notably, the CRNN is found to properly predict the stationary states consistent with the learned equation, despite never being observed at train time. A main limitation of the present approach is that the trained model is only applicable to Cahn-Hilliard flow and new training, together with an ad-hoc training set, is required should the evolution law be changed. Additionally,should a non-conservative phenomenon be considered, e.g. dendritic growth or grain coarsening, a suitable modification of the physics-inspired layer should be implemented. Due to the flexibility of NN approaches and successful applications to other microstructure evolution problems~\cite{Yang2021Patterns, Lanzoni2022PRM, peivaste2022machine, fan2024accelerate}, however, we expect that a suitably adapted version of the approach used in this work to be effective. On the other hand, if application to more challenging but related dynamical problems, such as evolution by surface diffusion, are to be tackled, the present model trained on Cahn-Hilliard flow could provide a good starting point for transfer learning procedures.

One of the main drawbacks of ML methods in physics is the possibility of extrapolation errors leading to inaccurate or even physically unrealistic predictions. In this respect, the present approach is shown to deliver close correspondence with the underlying physical driving force on a quantitative level, even when $\varphi$ values are no longer in one-to-one accordance. These characteristics, together with the long-time stability properties, pave the way to applications to more complex physical models and show the importance of physics-driven design for CRNN architectures.

\section*{Conflicts of interest}
The authors declare no conflicts of interest.

\section*{Data availability statement}
The data supporting the findings of this study are available upon reasonable request from the authors. Machine Learning code implementation is openly available at the following URL:~\url{https://github.com/dlanzo/CRANE}.

\begin{acknowledgments}
R.B., A.F and F.M. acknowledge financial support from ICSC – Centro Nazionale di Ricerca in High Performance Computing, Big Data and Quantum Computing, funded by European Union – NextGenerationEU.

D.L. acknowledges hardware support from Luca Lanzoni.
\end{acknowledgments}

\clearpage

\begin{appendix}

\section{Computational costs assessment for the NN method}
\label{sec::computational_costs}

\begin{figure}[ht]
    \centering
    \includegraphics[width=\columnwidth]{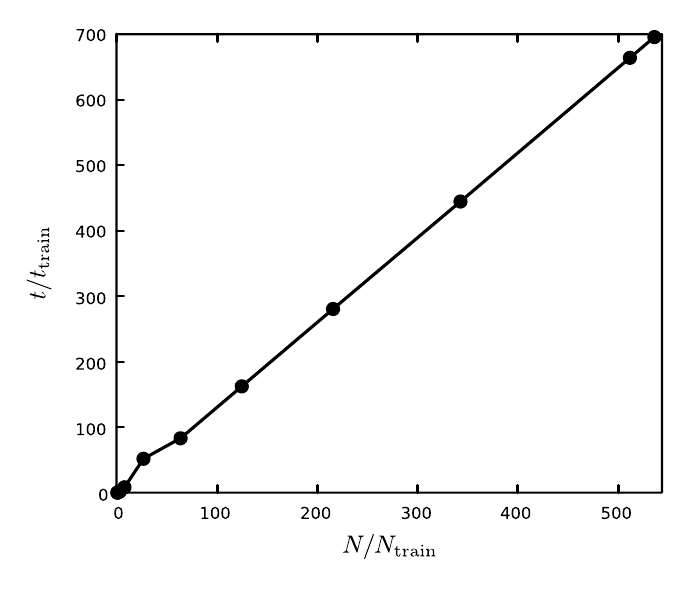}
    \caption{Computational cost for 1000 timestep prediction as a function of the number of collocation points in the domain. The number of collocation points is reported relative to the training set. $O(N)$ computational complexity can be observed. No additional optimization with respect to standard PyTorch implementation has been performed.}
    \label{fig::times}
\end{figure}

Here we report the scaling of the used algorithm with respect to domain size. Figure~\ref{fig::times} shows the wall time required to perform 1000 $\tau$ steps as a function of the total collocation points $N$ on a workstation using an Nvidia RTX A4000 GPU. Quantities are reported as a ratio with respect to the training set used in the main text. The computational cost is linear in the number of collocation points, as expected from the parallel implementation for the convolutional layers. Notably, this is more efficient than the $O( N\log N)$ scaling of pseudo-spectral schemes.

\section{Mesh rescaling effect on RMSE evaluation} \label{sec::downscaling_effect}

\begin{figure}[ht]
    \centering
    \includegraphics[width=\columnwidth]{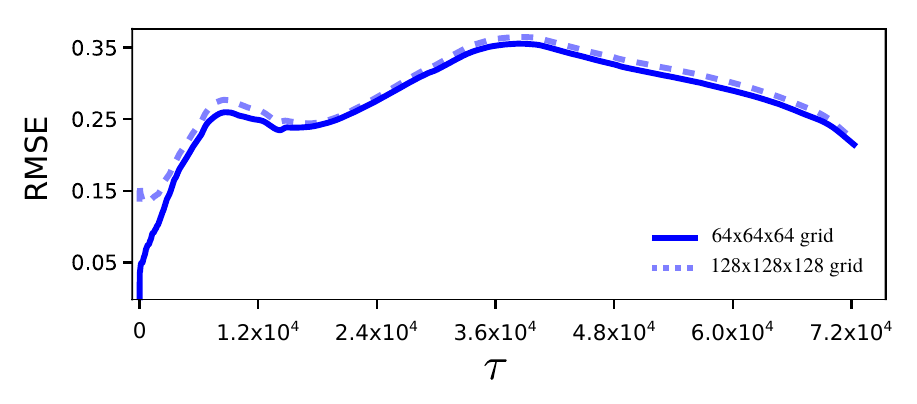}
    \caption{RMSE curve between the predicted and the ground truth $\varphi$ (rescaled on a $64 \times 64 \times 64$ grid) for the same evolution of Fig.~\ref{fig::phi_quantification}. The original curve on the $128 \times 128 \times 128$ grid is reported for comparison}
    \label{fig::rescaling_RMSE}
\end{figure}

We here report (Fig.~\ref{fig::rescaling_RMSE}) the RMSE plot of Fig.~\ref{fig::phi_quantification} as evaluated after downscaling the ground truth evolution on the same $64 \times 64 \times 64$ grid of collocation points used by the NN. Notice that the initial phases of the evolution now exhibit a lower RMSE value, due to the compensation of the resolution difference artifacts. In the subsequent phases of the evolution, however, values are only slightly affected, showing that the downscaling procedure is less relevant for coarser morphologies and does not significantly impact the predictive capabilities of the NN model. As the RMSE on the $64 \times 64 \times 64$ grid is smaller than the one evaluated on the $128 \times 128 \times 128$ one, we report only the latter on the main text as it provides a stricter error measure.

\section{Error measure} \label{sec::measure}

We here discuss the physical meaning of the $\bar{\delta}$ error measure reported in Eq.~\ref{eq::measure}. First, let us consider a phase field representation of a microstructure $\varphi(x)$, which implicitly defines the interface between the two phases through the $\varphi=0.5$ isoline. Suppose that the interface displaces with respect to its original position by a small distance $\delta(S)$. $S$ represents the intrinsic coordinates (locally) parallel to the interface. To leading order in $\delta$, the new phase field representation is given by:
\begin{equation}
    \hat{\varphi} (x) = \varphi(x) + \hat{n} \cdot \nabla{\varphi} (x) \delta(S)
\end{equation}
with $\hat{n} = \nabla{\varphi}/|\nabla{\varphi}|$ the surface normal. Then
\begin{equation}
    (\varphi - \hat{\varphi})^2 = \delta^2(S) |\nabla \varphi|^2
\end{equation}
and we define the mean squared interface displacement as
\begin{equation}
    \bar{\delta}^2 = \frac{\int_S \delta^2(S) dS}{A}
\end{equation}
where $A$ is the interface area and $\int_S ... dS$ is integration on the interface manifold.

Since both $(\varphi-\hat{\varphi})^2$ and $\nabla{\varphi}$ are localized near the interface, we can perform integrals on the whole domain $\int_\Omega ... dx $ involving these quantities using local parallel and normal coordinates $S$ and $\xi$ respectively. We then obtain that:
\begin{equation}
\begin{split}
    & \frac{\int_\Omega (\varphi - \hat{\varphi})^2 dx}{\int_\Omega |\nabla{\varphi}| dx} = \frac{\int_S \delta^2(S) dS \int_{\hat{n}} (\partial_\xi \varphi)^2 d\xi}{\int_S dS \int_{\hat{n}} (\partial_\xi \varphi)^2 d\xi} = \\
    & \frac{\int_S \delta^2(S) dS}{\int_S dS} = \bar{\delta}^2 ,
\end{split}
\end{equation}
where it is assumed in the second equality that the integral $\int  (\partial_\xi \phi)^2 d\xi$ does not depend on $S$, which is valid in the curvature $\ll$ interface thickness limit. Eq.~\ref{eq::measure} is recovered by square root.

\end{appendix}

\newpage

\setcounter{figure}{0}
\setcounter{section}{0}
\renewcommand*{\thefigure}{S\arabic{figure}}
\renewcommand*{\thesection}{S\arabic{section}}

\section{Performances assessment of NN performances if divergence layer is removed}
\label{sec::nodivergence_comparison}

We here report a comparison between the performances of the model considered in the main text and a variant that does not leverage the physics-inspired layer ("divergence" layer in Fig.~1(a) of the main text), showing that the first outperforms the latter one consistently, especially when very long sequences are considered. For the model version analyzed in this Section, the output layer is replaced by a logistic sigmoid activation function, following the approach used in 2D in Ref.~\cite{Lanzoni2022PRM}. Consistently with the same work, we also add to the loss function a term acting as a (weak) penalty for the non-conservation of $\varphi$ to provide a fair comparison. All other hyperparameters and the training set are identical to the $\varphi$-conserving model. In the following, we will refer to the model considered in the main text as the "conservative model" and the variant not implementing the divergence layer described above as the "non-conservative model".

\begin{figure}[ht]
    \centering
    \includegraphics[width=\columnwidth]{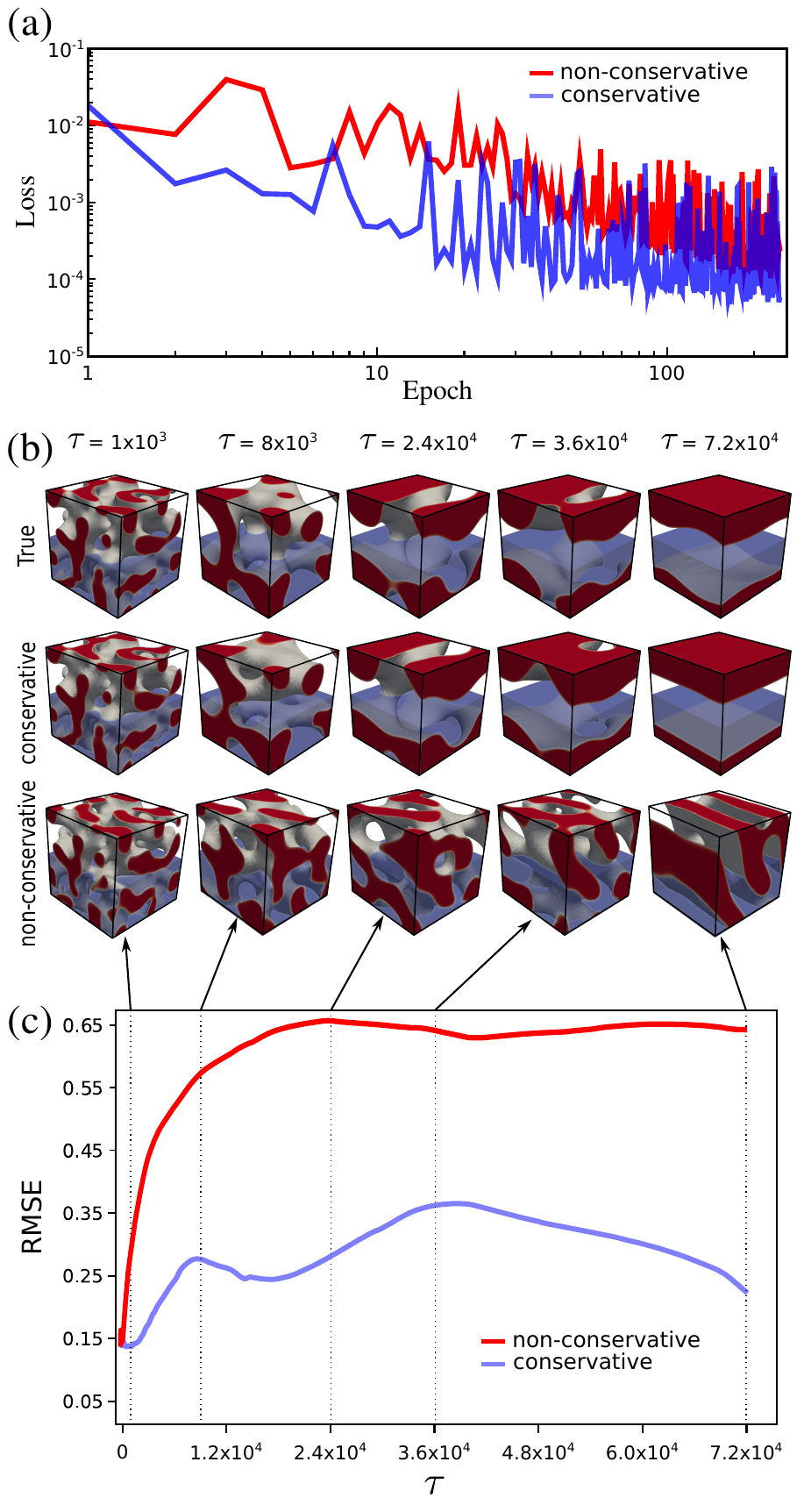}
    \caption{(a) Comparison between validation losses (calculated on the same set) for the conservative model discussed in the main text and a non-conservative one which does not implement the physics-inspired divergence layer. (b) Comparison between the finite difference ground truth, the conservative model prediction, and the non-conservative model one (evolution is the same as in Fig.~5 in the main text). (c) Comparison between the conservative and non-conservative model RMSE on the sequence of in panel (b).}
    \label{fig::nonconservative}
\end{figure}

Fig.~\ref{fig::nonconservative}(a) reports a comparison between the validation loss for the two models, calculated on the same set. A logarithmic scale is used to allow better inspection of small differences. Despite the oscillations, the conserving model consistently outperforms the non-conserving one.

Fig.~\ref{fig::nonconservative}(b) reports a comparison between the evolution obtained using the finite difference scheme and the prediction provided by both the conservative and non-conservative models. The initial condition is identical to the one of the evolution reported in Fig.~5(a) of the main text. Significant deviations are found in the sequence predicted by the non-conservative model in the late stages of the evolution, where a striking difference with respect to the expected layered configuration is found. It is also important to point out that error accumulation does not only affect the morphological prediction of the non-conservative model: while for short sequences in training only a fraction of percent in the deviation of the average value of $\varphi$ is observed, a high number of iterations such as the one reported in Fig.~\ref{fig::nonconservative}(b) leads to an $\approx 10\%$ deviation at $\tau = 7.2\times 10^4$. This contrasts with the conservative model, in which the phase field is conserved up to numerical precision.

Finally, we also report in Fig.~\ref{fig::nonconservative}(c) the RMSE curve with respect to the ground truth along the whole temporal sequence. The equivalent error measure for the conservative model, already present in Fig.~5(b) of the main text is also reported for comparison. While in the first $\approx 200 \tau$ the RMSE values are close, the non-conservative model error steeply increases when extrapolation with respect to the training set begins, saturating to a value almost twice the one observed for the conservative model. It can therefore be concluded that the use of the "divergence" layer in the CRNN structure is critical for accurate and physically consistent predictions in the long-time regimes.

\newpage

\bibliography{biblio}

%merlin.mbs apsrev4-1.bst 2010-07-25 4.21a (PWD, AO, DPC) hacked
%Control: key (0)
%Control: author (8) initials jnrlst
%Control: editor formatted (1) identically to author
%Control: production of article title (-1) disabled
%Control: page (0) single
%Control: year (1) truncated
%Control: production of eprint (0) enabled
\begin{thebibliography}{46}%
\makeatletter
\providecommand \@ifxundefined [1]{%
 \@ifx{#1\undefined}
}%
\providecommand \@ifnum [1]{%
 \ifnum #1\expandafter \@firstoftwo
 \else \expandafter \@secondoftwo
 \fi
}%
\providecommand \@ifx [1]{%
 \ifx #1\expandafter \@firstoftwo
 \else \expandafter \@secondoftwo
 \fi
}%
\providecommand \natexlab [1]{#1}%
\providecommand \enquote  [1]{``#1''}%
\providecommand \bibnamefont  [1]{#1}%
\providecommand \bibfnamefont [1]{#1}%
\providecommand \citenamefont [1]{#1}%
\providecommand \href@noop [0]{\@secondoftwo}%
\providecommand \href [0]{\begingroup \@sanitize@url \@href}%
\providecommand \@href[1]{\@@startlink{#1}\@@href}%
\providecommand \@@href[1]{\endgroup#1\@@endlink}%
\providecommand \@sanitize@url [0]{\catcode `\\12\catcode `\$12\catcode
  `\&12\catcode `\#12\catcode `\^12\catcode `\_12\catcode `\%12\relax}%
\providecommand \@@startlink[1]{}%
\providecommand \@@endlink[0]{}%
\providecommand \url  [0]{\begingroup\@sanitize@url \@url }%
\providecommand \@url [1]{\endgroup\@href {#1}{\urlprefix }}%
\providecommand \urlprefix  [0]{URL }%
\providecommand \Eprint [0]{\href }%
\providecommand \doibase [0]{http://dx.doi.org/}%
\providecommand \selectlanguage [0]{\@gobble}%
\providecommand \bibinfo  [0]{\@secondoftwo}%
\providecommand \bibfield  [0]{\@secondoftwo}%
\providecommand \translation [1]{[#1]}%
\providecommand \BibitemOpen [0]{}%
\providecommand \bibitemStop [0]{}%
\providecommand \bibitemNoStop [0]{.\EOS\space}%
\providecommand \EOS [0]{\spacefactor3000\relax}%
\providecommand \BibitemShut  [1]{\csname bibitem#1\endcsname}%
\let\auto@bib@innerbib\@empty
%</preamble>
\bibitem [{\citenamefont {Bishop}(2006)}]{bishop2006pattern}%
  \BibitemOpen
  \bibfield  {author} {\bibinfo {author} {\bibfnamefont {C.~M.}\ \bibnamefont
  {Bishop}},\ }\href@noop {} {\emph {\bibinfo {title} {Pattern recognition and
  machine learning}}}\ (\bibinfo  {publisher} {Springer New York, NY},\
  \bibinfo {year} {2006})\BibitemShut {NoStop}%
\bibitem [{\citenamefont {Goodfellow}\ \emph {et~al.}(2016)\citenamefont
  {Goodfellow}, \citenamefont {Bengio},\ and\ \citenamefont
  {Courville}}]{goodfellowdeep2016}%
  \BibitemOpen
  \bibfield  {author} {\bibinfo {author} {\bibfnamefont {I.}~\bibnamefont
  {Goodfellow}}, \bibinfo {author} {\bibfnamefont {Y.}~\bibnamefont {Bengio}},
  \ and\ \bibinfo {author} {\bibfnamefont {A.}~\bibnamefont {Courville}},\
  }\href@noop {} {\emph {\bibinfo {title} {Deep Learning}}}\ (\bibinfo
  {publisher} {MIT Press},\ \bibinfo {year} {2016})\ \bibinfo {note}
  {\url{http://www.deeplearningbook.org}}\BibitemShut {NoStop}%
\bibitem [{\citenamefont {Butler}\ \emph {et~al.}(2018)\citenamefont {Butler},
  \citenamefont {Davies}, \citenamefont {Cartwright}, \citenamefont {Isayev},\
  and\ \citenamefont {Walsh}}]{butler2018machine}%
  \BibitemOpen
  \bibfield  {author} {\bibinfo {author} {\bibfnamefont {K.~T.}\ \bibnamefont
  {Butler}}, \bibinfo {author} {\bibfnamefont {D.~W.}\ \bibnamefont {Davies}},
  \bibinfo {author} {\bibfnamefont {H.}~\bibnamefont {Cartwright}}, \bibinfo
  {author} {\bibfnamefont {O.}~\bibnamefont {Isayev}}, \ and\ \bibinfo {author}
  {\bibfnamefont {A.}~\bibnamefont {Walsh}},\ }\href@noop {} {\bibfield
  {journal} {\bibinfo  {journal} {Nature}\ }\textbf {\bibinfo {volume} {559}},\
  \bibinfo {pages} {547} (\bibinfo {year} {2018})}\BibitemShut {NoStop}%
\bibitem [{\citenamefont {Mehta}\ \emph {et~al.}(2019)\citenamefont {Mehta},
  \citenamefont {Bukov}, \citenamefont {Wang}, \citenamefont {Day},
  \citenamefont {Richardson}, \citenamefont {Fisher},\ and\ \citenamefont
  {Schwab}}]{Mehta2019PhysRep}%
  \BibitemOpen
  \bibfield  {author} {\bibinfo {author} {\bibfnamefont {P.}~\bibnamefont
  {Mehta}}, \bibinfo {author} {\bibfnamefont {M.}~\bibnamefont {Bukov}},
  \bibinfo {author} {\bibfnamefont {C.~H.}\ \bibnamefont {Wang}}, \bibinfo
  {author} {\bibfnamefont {A.~G.}\ \bibnamefont {Day}}, \bibinfo {author}
  {\bibfnamefont {C.}~\bibnamefont {Richardson}}, \bibinfo {author}
  {\bibfnamefont {C.~K.}\ \bibnamefont {Fisher}}, \ and\ \bibinfo {author}
  {\bibfnamefont {D.~J.}\ \bibnamefont {Schwab}},\ }\href {\doibase
  10.1016/j.physrep.2019.03.001} {\bibfield  {journal} {\bibinfo  {journal}
  {Physics Reports}\ }\textbf {\bibinfo {volume} {810}},\ \bibinfo {pages} {1}
  (\bibinfo {year} {2019})}\BibitemShut {NoStop}%
\bibitem [{\citenamefont {Bedolla}\ \emph {et~al.}(2020)\citenamefont
  {Bedolla}, \citenamefont {Padierna},\ and\ \citenamefont
  {Castaneda-Priego}}]{bedolla2020machine}%
  \BibitemOpen
  \bibfield  {author} {\bibinfo {author} {\bibfnamefont {E.}~\bibnamefont
  {Bedolla}}, \bibinfo {author} {\bibfnamefont {L.~C.}\ \bibnamefont
  {Padierna}}, \ and\ \bibinfo {author} {\bibfnamefont {R.}~\bibnamefont
  {Castaneda-Priego}},\ }\href@noop {} {\bibfield  {journal} {\bibinfo
  {journal} {Journal of Physics: Condensed Matter}\ }\textbf {\bibinfo {volume}
  {33}},\ \bibinfo {pages} {053001} (\bibinfo {year} {2020})}\BibitemShut
  {NoStop}%
\bibitem [{\citenamefont {Nguyen}\ \emph {et~al.}(2024)\citenamefont {Nguyen},
  \citenamefont {Potapenko}, \citenamefont {Demirci}, \citenamefont {Govind},
  \citenamefont {Bompas},\ and\ \citenamefont {Sandfeld}}]{NGUYEN2024100544}%
  \BibitemOpen
  \bibfield  {author} {\bibinfo {author} {\bibfnamefont {B.~D.}\ \bibnamefont
  {Nguyen}}, \bibinfo {author} {\bibfnamefont {P.}~\bibnamefont {Potapenko}},
  \bibinfo {author} {\bibfnamefont {A.}~\bibnamefont {Demirci}}, \bibinfo
  {author} {\bibfnamefont {K.}~\bibnamefont {Govind}}, \bibinfo {author}
  {\bibfnamefont {S.}~\bibnamefont {Bompas}}, \ and\ \bibinfo {author}
  {\bibfnamefont {S.}~\bibnamefont {Sandfeld}},\ }\href {\doibase
  https://doi.org/10.1016/j.mlwa.2024.100544} {\bibfield  {journal} {\bibinfo
  {journal} {Machine Learning with Applications}\ }\textbf {\bibinfo {volume}
  {16}},\ \bibinfo {pages} {100544} (\bibinfo {year} {2024})}\BibitemShut
  {NoStop}%
\bibitem [{\citenamefont {Bart{\'{o}}k}\ \emph {et~al.}(2010)\citenamefont
  {Bart{\'{o}}k}, \citenamefont {Payne}, \citenamefont {Kondor},\ and\
  \citenamefont {Cs{\'{a}}nyi}}]{Bartok2010PRL}%
  \BibitemOpen
  \bibfield  {author} {\bibinfo {author} {\bibfnamefont {A.~P.}\ \bibnamefont
  {Bart{\'{o}}k}}, \bibinfo {author} {\bibfnamefont {M.~P.}\ \bibnamefont
  {Payne}}, \bibinfo {author} {\bibfnamefont {R.}~\bibnamefont {Kondor}}, \
  and\ \bibinfo {author} {\bibfnamefont {G.}~\bibnamefont {Cs{\'{a}}nyi}},\
  }\href {\doibase 10.1103/PhysRevLett.104.136403} {\bibfield  {journal}
  {\bibinfo  {journal} {Physical Review Letters}\ }\textbf {\bibinfo {volume}
  {104}},\ \bibinfo {pages} {136403} (\bibinfo {year} {2010})}\BibitemShut
  {NoStop}%
\bibitem [{\citenamefont {Kocer}\ \emph {et~al.}(2022)\citenamefont {Kocer},
  \citenamefont {Ko},\ and\ \citenamefont {Behler}}]{Kocer2022review}%
  \BibitemOpen
  \bibfield  {author} {\bibinfo {author} {\bibfnamefont {E.}~\bibnamefont
  {Kocer}}, \bibinfo {author} {\bibfnamefont {T.~W.}\ \bibnamefont {Ko}}, \
  and\ \bibinfo {author} {\bibfnamefont {J.}~\bibnamefont {Behler}},\ }\href
  {\doibase 10.1146/annurev-physchem-082720-034254} {\bibfield  {journal}
  {\bibinfo  {journal} {Annual Review of Physical Chemistry}\ }\textbf
  {\bibinfo {volume} {73}},\ \bibinfo {pages} {163} (\bibinfo {year}
  {2022})}\BibitemShut {NoStop}%
\bibitem [{\citenamefont {Kim}\ \emph {et~al.}(2019)\citenamefont {Kim},
  \citenamefont {Azevedo}, \citenamefont {Thuerey}, \citenamefont {Kim},
  \citenamefont {Gross},\ and\ \citenamefont {Solenthaler}}]{Kim2019CGF}%
  \BibitemOpen
  \bibfield  {author} {\bibinfo {author} {\bibfnamefont {B.}~\bibnamefont
  {Kim}}, \bibinfo {author} {\bibfnamefont {V.~C.}\ \bibnamefont {Azevedo}},
  \bibinfo {author} {\bibfnamefont {N.}~\bibnamefont {Thuerey}}, \bibinfo
  {author} {\bibfnamefont {T.}~\bibnamefont {Kim}}, \bibinfo {author}
  {\bibfnamefont {M.}~\bibnamefont {Gross}}, \ and\ \bibinfo {author}
  {\bibfnamefont {B.}~\bibnamefont {Solenthaler}},\ }\href {\doibase
  10.1111/cgf.13619} {\bibfield  {journal} {\bibinfo  {journal} {Computer
  Graphics Forum}\ }\textbf {\bibinfo {volume} {38}},\ \bibinfo {pages} {59}
  (\bibinfo {year} {2019})},\ \Eprint {http://arxiv.org/abs/1806.02071}
  {1806.02071} \BibitemShut {NoStop}%
\bibitem [{\citenamefont {Fulton}\ \emph {et~al.}(2019)\citenamefont {Fulton},
  \citenamefont {Modi}, \citenamefont {Duvenaud}, \citenamefont {Levin},\ and\
  \citenamefont {Jacobson}}]{Fulton2019CGF}%
  \BibitemOpen
  \bibfield  {author} {\bibinfo {author} {\bibfnamefont {L.}~\bibnamefont
  {Fulton}}, \bibinfo {author} {\bibfnamefont {V.}~\bibnamefont {Modi}},
  \bibinfo {author} {\bibfnamefont {D.}~\bibnamefont {Duvenaud}}, \bibinfo
  {author} {\bibfnamefont {D.~I.}\ \bibnamefont {Levin}}, \ and\ \bibinfo
  {author} {\bibfnamefont {A.}~\bibnamefont {Jacobson}},\ }\href {\doibase
  10.1111/cgf.13645} {\bibfield  {journal} {\bibinfo  {journal} {Computer
  Graphics Forum}\ }\textbf {\bibinfo {volume} {38}},\ \bibinfo {pages} {379}
  (\bibinfo {year} {2019})}\BibitemShut {NoStop}%
\bibitem [{\citenamefont {Zhang}\ and\ \citenamefont
  {Garikipati}(2020)}]{Zhang2020CMAME}%
  \BibitemOpen
  \bibfield  {author} {\bibinfo {author} {\bibfnamefont {X.}~\bibnamefont
  {Zhang}}\ and\ \bibinfo {author} {\bibfnamefont {K.}~\bibnamefont
  {Garikipati}},\ }\href {\doibase 10.1016/j.cma.2020.113362} {\bibfield
  {journal} {\bibinfo  {journal} {Computer Methods in Applied Mechanics and
  Engineering}\ }\textbf {\bibinfo {volume} {372}},\ \bibinfo {pages} {1}
  (\bibinfo {year} {2020})},\ \Eprint {http://arxiv.org/abs/2001.01575}
  {arXiv:2001.01575} \BibitemShut {NoStop}%
\bibitem [{\citenamefont {{Montes de Oca Zapiain}}\ \emph
  {et~al.}(2021)\citenamefont {{Montes de Oca Zapiain}}, \citenamefont
  {Stewart},\ and\ \citenamefont {Dingreville}}]{MontesdeOcaZapiain2021npj}%
  \BibitemOpen
  \bibfield  {author} {\bibinfo {author} {\bibfnamefont {D.}~\bibnamefont
  {{Montes de Oca Zapiain}}}, \bibinfo {author} {\bibfnamefont {J.~A.}\
  \bibnamefont {Stewart}}, \ and\ \bibinfo {author} {\bibfnamefont
  {R.}~\bibnamefont {Dingreville}},\ }\href {\doibase
  10.1038/s41524-020-00471-8} {\bibfield  {journal} {\bibinfo  {journal} {npj
  Computational Materials}\ }\textbf {\bibinfo {volume} {7}},\ \bibinfo {pages}
  {1} (\bibinfo {year} {2021})}\BibitemShut {NoStop}%
\bibitem [{\citenamefont {Yang}\ \emph {et~al.}(2021)\citenamefont {Yang},
  \citenamefont {Cao}, \citenamefont {Zhang}, \citenamefont {Fan},
  \citenamefont {Tang}, \citenamefont {Aberg}, \citenamefont {Sadigh},\ and\
  \citenamefont {Zhou}}]{Yang2021Patterns}%
  \BibitemOpen
  \bibfield  {author} {\bibinfo {author} {\bibfnamefont {K.}~\bibnamefont
  {Yang}}, \bibinfo {author} {\bibfnamefont {Y.}~\bibnamefont {Cao}}, \bibinfo
  {author} {\bibfnamefont {Y.}~\bibnamefont {Zhang}}, \bibinfo {author}
  {\bibfnamefont {S.}~\bibnamefont {Fan}}, \bibinfo {author} {\bibfnamefont
  {M.}~\bibnamefont {Tang}}, \bibinfo {author} {\bibfnamefont {D.}~\bibnamefont
  {Aberg}}, \bibinfo {author} {\bibfnamefont {B.}~\bibnamefont {Sadigh}}, \
  and\ \bibinfo {author} {\bibfnamefont {F.}~\bibnamefont {Zhou}},\ }\href
  {\doibase 10.1016/j.patter.2021.100243} {\bibfield  {journal} {\bibinfo
  {journal} {Patterns}\ }\textbf {\bibinfo {volume} {2}},\ \bibinfo {pages}
  {100243} (\bibinfo {year} {2021})}\BibitemShut {NoStop}%
\bibitem [{\citenamefont {Lanzoni}\ \emph {et~al.}(2022)\citenamefont
  {Lanzoni}, \citenamefont {Albani}, \citenamefont {Bergamaschini},\ and\
  \citenamefont {Montalenti}}]{Lanzoni2022PRM}%
  \BibitemOpen
  \bibfield  {author} {\bibinfo {author} {\bibfnamefont {D.}~\bibnamefont
  {Lanzoni}}, \bibinfo {author} {\bibfnamefont {M.}~\bibnamefont {Albani}},
  \bibinfo {author} {\bibfnamefont {R.}~\bibnamefont {Bergamaschini}}, \ and\
  \bibinfo {author} {\bibfnamefont {F.}~\bibnamefont {Montalenti}},\ }\href
  {\doibase 10.1103/PhysRevMaterials.6.103801} {\bibfield  {journal} {\bibinfo
  {journal} {Physical Review Materials}\ }\textbf {\bibinfo {volume} {6}},\
  \bibinfo {pages} {103801} (\bibinfo {year} {2022})}\BibitemShut {NoStop}%
\bibitem [{\citenamefont {Martín-Encinar}\ \emph {et~al.}(2024)\citenamefont
  {Martín-Encinar}, \citenamefont {Lanzoni}, \citenamefont {Fantasia},
  \citenamefont {Rovaris}, \citenamefont {Bergamaschini},\ and\ \citenamefont
  {Montalenti}}]{LUIS}%
  \BibitemOpen
  \bibfield  {author} {\bibinfo {author} {\bibfnamefont {L.}~\bibnamefont
  {Martín-Encinar}}, \bibinfo {author} {\bibfnamefont {D.}~\bibnamefont
  {Lanzoni}}, \bibinfo {author} {\bibfnamefont {A.}~\bibnamefont {Fantasia}},
  \bibinfo {author} {\bibfnamefont {F.}~\bibnamefont {Rovaris}}, \bibinfo
  {author} {\bibfnamefont {R.}~\bibnamefont {Bergamaschini}}, \ and\ \bibinfo
  {author} {\bibfnamefont {F.}~\bibnamefont {Montalenti}},\ }\href
  {https://arxiv.org/abs/2405.03049} {\enquote {\bibinfo {title} {Quantitative
  analysis of the prediction performance of a convolutional neural network
  evaluating the surface elastic energy of a strained film},}\ } (\bibinfo
  {year} {2024}),\ \Eprint {http://arxiv.org/abs/2405.03049} {arXiv:2405.03049
  [physics.comp-ph]} \BibitemShut {NoStop}%
\bibitem [{\citenamefont {Raissi}\ \emph {et~al.}(2019)\citenamefont {Raissi},
  \citenamefont {Perdikaris},\ and\ \citenamefont
  {Karniadakis}}]{RAISSI2019PINN}%
  \BibitemOpen
  \bibfield  {author} {\bibinfo {author} {\bibfnamefont {M.}~\bibnamefont
  {Raissi}}, \bibinfo {author} {\bibfnamefont {P.}~\bibnamefont {Perdikaris}},
  \ and\ \bibinfo {author} {\bibfnamefont {G.}~\bibnamefont {Karniadakis}},\
  }\href {\doibase https://doi.org/10.1016/j.jcp.2018.10.045} {\bibfield
  {journal} {\bibinfo  {journal} {Journal of Computational Physics}\ }\textbf
  {\bibinfo {volume} {378}},\ \bibinfo {pages} {686} (\bibinfo {year}
  {2019})}\BibitemShut {NoStop}%
\bibitem [{\citenamefont {Bretin}\ \emph {et~al.}(2022)\citenamefont {Bretin},
  \citenamefont {Denis}, \citenamefont {Masnou},\ and\ \citenamefont
  {Terii}}]{Bretin2022JCompPhys}%
  \BibitemOpen
  \bibfield  {author} {\bibinfo {author} {\bibfnamefont {E.}~\bibnamefont
  {Bretin}}, \bibinfo {author} {\bibfnamefont {R.}~\bibnamefont {Denis}},
  \bibinfo {author} {\bibfnamefont {S.}~\bibnamefont {Masnou}}, \ and\ \bibinfo
  {author} {\bibfnamefont {G.}~\bibnamefont {Terii}},\ }\href {\doibase
  https://doi.org/10.1016/j.jcp.2022.111579} {\bibfield  {journal} {\bibinfo
  {journal} {Journal of Computational Physics}\ }\textbf {\bibinfo {volume}
  {470}},\ \bibinfo {pages} {111579} (\bibinfo {year} {2022})}\BibitemShut
  {NoStop}%
\bibitem [{\citenamefont {Peivaste}\ \emph {et~al.}(2022)\citenamefont
  {Peivaste}, \citenamefont {Siboni}, \citenamefont {Alahyarizadeh},
  \citenamefont {Ghaderi}, \citenamefont {Svendsen}, \citenamefont {Raabe},\
  and\ \citenamefont {Mianroodi}}]{peivaste2022machine}%
  \BibitemOpen
  \bibfield  {author} {\bibinfo {author} {\bibfnamefont {I.}~\bibnamefont
  {Peivaste}}, \bibinfo {author} {\bibfnamefont {N.~H.}\ \bibnamefont
  {Siboni}}, \bibinfo {author} {\bibfnamefont {G.}~\bibnamefont
  {Alahyarizadeh}}, \bibinfo {author} {\bibfnamefont {R.}~\bibnamefont
  {Ghaderi}}, \bibinfo {author} {\bibfnamefont {B.}~\bibnamefont {Svendsen}},
  \bibinfo {author} {\bibfnamefont {D.}~\bibnamefont {Raabe}}, \ and\ \bibinfo
  {author} {\bibfnamefont {J.~R.}\ \bibnamefont {Mianroodi}},\ }\href@noop {}
  {\bibfield  {journal} {\bibinfo  {journal} {Computational Materials Science}\
  }\textbf {\bibinfo {volume} {214}},\ \bibinfo {pages} {111750} (\bibinfo
  {year} {2022})}\BibitemShut {NoStop}%
\bibitem [{\citenamefont {Fan}\ \emph {et~al.}(2024)\citenamefont {Fan},
  \citenamefont {Hitt}, \citenamefont {Tang}, \citenamefont {Sadigh},\ and\
  \citenamefont {Zhou}}]{fan2024accelerate}%
  \BibitemOpen
  \bibfield  {author} {\bibinfo {author} {\bibfnamefont {S.}~\bibnamefont
  {Fan}}, \bibinfo {author} {\bibfnamefont {A.~L.}\ \bibnamefont {Hitt}},
  \bibinfo {author} {\bibfnamefont {M.}~\bibnamefont {Tang}}, \bibinfo {author}
  {\bibfnamefont {B.}~\bibnamefont {Sadigh}}, \ and\ \bibinfo {author}
  {\bibfnamefont {F.}~\bibnamefont {Zhou}},\ }\href@noop {} {\bibfield
  {journal} {\bibinfo  {journal} {Machine Learning: Science and Technology}\
  }\textbf {\bibinfo {volume} {5}},\ \bibinfo {pages} {025027} (\bibinfo {year}
  {2024})}\BibitemShut {NoStop}%
\bibitem [{\citenamefont {Langer}(1971)}]{Langer1971AnnPhy}%
  \BibitemOpen
  \bibfield  {author} {\bibinfo {author} {\bibfnamefont {J.}~\bibnamefont
  {Langer}},\ }\href {\doibase https://doi.org/10.1016/0003-4916(71)90162-X}
  {\bibfield  {journal} {\bibinfo  {journal} {Annals of Physics}\ }\textbf
  {\bibinfo {volume} {65}},\ \bibinfo {pages} {53} (\bibinfo {year}
  {1971})}\BibitemShut {NoStop}%
\bibitem [{\citenamefont {Kwon}\ \emph {et~al.}(2007)\citenamefont {Kwon},
  \citenamefont {Thornton},\ and\ \citenamefont {Voorhees}}]{Kwon2007PRE}%
  \BibitemOpen
  \bibfield  {author} {\bibinfo {author} {\bibfnamefont {Y.}~\bibnamefont
  {Kwon}}, \bibinfo {author} {\bibfnamefont {K.}~\bibnamefont {Thornton}}, \
  and\ \bibinfo {author} {\bibfnamefont {P.~W.}\ \bibnamefont {Voorhees}},\
  }\href {\doibase 10.1103/PhysRevE.75.021120} {\bibfield  {journal} {\bibinfo
  {journal} {Phys. Rev. E}\ }\textbf {\bibinfo {volume} {75}},\ \bibinfo
  {pages} {021120} (\bibinfo {year} {2007})}\BibitemShut {NoStop}%
\bibitem [{\citenamefont {Andrews}\ \emph {et~al.}(2020)\citenamefont
  {Andrews}, \citenamefont {Elder}, \citenamefont {Voorhees},\ and\
  \citenamefont {Thornton}}]{Andrews2020PRM}%
  \BibitemOpen
  \bibfield  {author} {\bibinfo {author} {\bibfnamefont {W.~B.}\ \bibnamefont
  {Andrews}}, \bibinfo {author} {\bibfnamefont {K.~L.}\ \bibnamefont {Elder}},
  \bibinfo {author} {\bibfnamefont {P.~W.}\ \bibnamefont {Voorhees}}, \ and\
  \bibinfo {author} {\bibfnamefont {K.}~\bibnamefont {Thornton}},\ }\href
  {\doibase 10.1103/PhysRevMaterials.4.103401} {\bibfield  {journal} {\bibinfo
  {journal} {Physical Review Materials}\ }\textbf {\bibinfo {volume} {4}},\
  \bibinfo {pages} {1} (\bibinfo {year} {2020})}\BibitemShut {NoStop}%
\bibitem [{\citenamefont {Jinnai}\ \emph {et~al.}(2000)\citenamefont {Jinnai},
  \citenamefont {Nishikawa}, \citenamefont {Morimoto}, \citenamefont {Koga},\
  and\ \citenamefont {Hashimoto}}]{jinnai2000geometrical}%
  \BibitemOpen
  \bibfield  {author} {\bibinfo {author} {\bibfnamefont {H.}~\bibnamefont
  {Jinnai}}, \bibinfo {author} {\bibfnamefont {Y.}~\bibnamefont {Nishikawa}},
  \bibinfo {author} {\bibfnamefont {H.}~\bibnamefont {Morimoto}}, \bibinfo
  {author} {\bibfnamefont {T.}~\bibnamefont {Koga}}, \ and\ \bibinfo {author}
  {\bibfnamefont {T.}~\bibnamefont {Hashimoto}},\ }\href@noop {} {\bibfield
  {journal} {\bibinfo  {journal} {Langmuir}\ }\textbf {\bibinfo {volume}
  {16}},\ \bibinfo {pages} {4380} (\bibinfo {year} {2000})}\BibitemShut
  {NoStop}%
\bibitem [{\citenamefont {Li}\ \emph {et~al.}(2009)\citenamefont {Li},
  \citenamefont {Lowengrub}, \citenamefont {R{\"{a}}tz},\ and\ \citenamefont
  {Voigt}}]{Li2009CommunComput}%
  \BibitemOpen
  \bibfield  {author} {\bibinfo {author} {\bibfnamefont {B.}~\bibnamefont
  {Li}}, \bibinfo {author} {\bibfnamefont {J.}~\bibnamefont {Lowengrub}},
  \bibinfo {author} {\bibfnamefont {A.}~\bibnamefont {R{\"{a}}tz}}, \ and\
  \bibinfo {author} {\bibfnamefont {A.}~\bibnamefont {Voigt}},\ }\href@noop {}
  {\bibfield  {journal} {\bibinfo  {journal} {Communications in Computational
  Physics}\ }\textbf {\bibinfo {volume} {6}},\ \bibinfo {pages} {433} (\bibinfo
  {year} {2009})}\BibitemShut {NoStop}%
\bibitem [{\citenamefont {Provatas}\ and\ \citenamefont
  {Elder}(2011)}]{provatas2011phase}%
  \BibitemOpen
  \bibfield  {author} {\bibinfo {author} {\bibfnamefont {N.}~\bibnamefont
  {Provatas}}\ and\ \bibinfo {author} {\bibfnamefont {K.}~\bibnamefont
  {Elder}},\ }\href@noop {} {\emph {\bibinfo {title} {Phase-field methods in
  materials science and engineering}}}\ (\bibinfo  {publisher} {John Wiley \&
  Sons},\ \bibinfo {year} {2011})\BibitemShut {NoStop}%
\bibitem [{\citenamefont {Chen}(2002)}]{chen2002phase}%
  \BibitemOpen
  \bibfield  {author} {\bibinfo {author} {\bibfnamefont {L.-Q.}\ \bibnamefont
  {Chen}},\ }\href@noop {} {\bibfield  {journal} {\bibinfo  {journal} {Annual
  review of materials research}\ }\textbf {\bibinfo {volume} {32}},\ \bibinfo
  {pages} {113} (\bibinfo {year} {2002})}\BibitemShut {NoStop}%
\bibitem [{\citenamefont {Boettinger}\ \emph {et~al.}(2002)\citenamefont
  {Boettinger}, \citenamefont {Warren}, \citenamefont {Beckermann},\ and\
  \citenamefont {Karma}}]{boettinger2002phase}%
  \BibitemOpen
  \bibfield  {author} {\bibinfo {author} {\bibfnamefont {W.~J.}\ \bibnamefont
  {Boettinger}}, \bibinfo {author} {\bibfnamefont {J.~A.}\ \bibnamefont
  {Warren}}, \bibinfo {author} {\bibfnamefont {C.}~\bibnamefont {Beckermann}},
  \ and\ \bibinfo {author} {\bibfnamefont {A.}~\bibnamefont {Karma}},\
  }\href@noop {} {\bibfield  {journal} {\bibinfo  {journal} {Annual review of
  materials research}\ }\textbf {\bibinfo {volume} {32}},\ \bibinfo {pages}
  {163} (\bibinfo {year} {2002})}\BibitemShut {NoStop}%
\bibitem [{\citenamefont {Albani}\ \emph {et~al.}(2016)\citenamefont {Albani},
  \citenamefont {Bergamaschini},\ and\ \citenamefont {Montalenti}}]{albani}%
  \BibitemOpen
  \bibfield  {author} {\bibinfo {author} {\bibfnamefont {M.}~\bibnamefont
  {Albani}}, \bibinfo {author} {\bibfnamefont {R.}~\bibnamefont
  {Bergamaschini}}, \ and\ \bibinfo {author} {\bibfnamefont {F.}~\bibnamefont
  {Montalenti}},\ }\href {\doibase 10.1103/PhysRevB.94.075303} {\bibfield
  {journal} {\bibinfo  {journal} {Phys. Rev. B}\ }\textbf {\bibinfo {volume}
  {94}},\ \bibinfo {pages} {075303} (\bibinfo {year} {2016})}\BibitemShut
  {NoStop}%
\bibitem [{\citenamefont {Wang}\ \emph {et~al.}(2023)\citenamefont {Wang},
  \citenamefont {Dabaja}, \citenamefont {Chen},\ and\ \citenamefont
  {Banu}}]{wang2023machine}%
  \BibitemOpen
  \bibfield  {author} {\bibinfo {author} {\bibfnamefont {Z.}~\bibnamefont
  {Wang}}, \bibinfo {author} {\bibfnamefont {R.}~\bibnamefont {Dabaja}},
  \bibinfo {author} {\bibfnamefont {L.}~\bibnamefont {Chen}}, \ and\ \bibinfo
  {author} {\bibfnamefont {M.}~\bibnamefont {Banu}},\ }\href@noop {} {\bibfield
   {journal} {\bibinfo  {journal} {Scientific Reports}\ }\textbf {\bibinfo
  {volume} {13}},\ \bibinfo {pages} {5414} (\bibinfo {year}
  {2023})}\BibitemShut {NoStop}%
\bibitem [{\citenamefont {Cahn}\ and\ \citenamefont
  {Hilliard}(1958)}]{cahn1958free}%
  \BibitemOpen
  \bibfield  {author} {\bibinfo {author} {\bibfnamefont {J.~W.}\ \bibnamefont
  {Cahn}}\ and\ \bibinfo {author} {\bibfnamefont {J.~E.}\ \bibnamefont
  {Hilliard}},\ }\href@noop {} {\bibfield  {journal} {\bibinfo  {journal} {The
  Journal of chemical physics}\ }\textbf {\bibinfo {volume} {28}},\ \bibinfo
  {pages} {258} (\bibinfo {year} {1958})}\BibitemShut {NoStop}%
\bibitem [{\citenamefont {Cahn}(1965)}]{cahn1965JCP}%
  \BibitemOpen
  \bibfield  {author} {\bibinfo {author} {\bibfnamefont {J.~W.}\ \bibnamefont
  {Cahn}},\ }\href {\doibase 10.1063/1.1695731} {\bibfield  {journal} {\bibinfo
   {journal} {The Journal of Chemical Physics}\ }\textbf {\bibinfo {volume}
  {42}},\ \bibinfo {pages} {93} (\bibinfo {year} {1965})}\BibitemShut {NoStop}%
\bibitem [{\citenamefont {Kumar}\ \emph {et~al.}(2020)\citenamefont {Kumar},
  \citenamefont {Tan}, \citenamefont {Zheng},\ and\ \citenamefont
  {Kochmann}}]{kumar2020inverse}%
  \BibitemOpen
  \bibfield  {author} {\bibinfo {author} {\bibfnamefont {S.}~\bibnamefont
  {Kumar}}, \bibinfo {author} {\bibfnamefont {S.}~\bibnamefont {Tan}}, \bibinfo
  {author} {\bibfnamefont {L.}~\bibnamefont {Zheng}}, \ and\ \bibinfo {author}
  {\bibfnamefont {D.~M.}\ \bibnamefont {Kochmann}},\ }\href@noop {} {\bibfield
  {journal} {\bibinfo  {journal} {npj Computational Materials}\ }\textbf
  {\bibinfo {volume} {6}},\ \bibinfo {pages} {73} (\bibinfo {year}
  {2020})}\BibitemShut {NoStop}%
\bibitem [{\citenamefont {Perlin}(1985)}]{Perlin1985287}%
  \BibitemOpen
  \bibfield  {author} {\bibinfo {author} {\bibfnamefont {K.}~\bibnamefont
  {Perlin}},\ }\href {\doibase 10.1145/325165.325247} {\bibfield  {journal}
  {\bibinfo  {journal} {SIGGRAPH Computer Graphics}\ }\textbf {\bibinfo
  {volume} {19}},\ \bibinfo {pages} {287 – 296} (\bibinfo {year}
  {1985})}\BibitemShut {NoStop}%
\bibitem [{Pyt(2023)}]{PythonPerlinNoise}%
  \BibitemOpen
  \href@noop {} {\enquote {\bibinfo {title} {Python implementation for perlin
  noise},}\ }\bibinfo {howpublished}
  {\url{https://pypi.org/project/perlin-noise/ }} (\bibinfo {year}
  {2023})\BibitemShut {NoStop}%
\bibitem [{\citenamefont {Paszke}\ \emph {et~al.}(2019)\citenamefont {Paszke},
  \citenamefont {Gross}, \citenamefont {Massa}, \citenamefont {Lerer},
  \citenamefont {Bradbury}, \citenamefont {Chanan}, \citenamefont {Killeen},
  \citenamefont {Lin}, \citenamefont {Gimelshein}, \citenamefont {Antiga},
  \citenamefont {Desmaison}, \citenamefont {Köpf}, \citenamefont {Yang},
  \citenamefont {DeVito}, \citenamefont {Raison}, \citenamefont {Tejani},
  \citenamefont {Chilamkurthy}, \citenamefont {Steiner}, \citenamefont {Fang},
  \citenamefont {Bai},\ and\ \citenamefont {Chintala}}]{paszkepytorch2019}%
  \BibitemOpen
  \bibfield  {author} {\bibinfo {author} {\bibfnamefont {A.}~\bibnamefont
  {Paszke}}, \bibinfo {author} {\bibfnamefont {S.}~\bibnamefont {Gross}},
  \bibinfo {author} {\bibfnamefont {F.}~\bibnamefont {Massa}}, \bibinfo
  {author} {\bibfnamefont {A.}~\bibnamefont {Lerer}}, \bibinfo {author}
  {\bibfnamefont {J.}~\bibnamefont {Bradbury}}, \bibinfo {author}
  {\bibfnamefont {G.}~\bibnamefont {Chanan}}, \bibinfo {author} {\bibfnamefont
  {T.}~\bibnamefont {Killeen}}, \bibinfo {author} {\bibfnamefont
  {Z.}~\bibnamefont {Lin}}, \bibinfo {author} {\bibfnamefont {N.}~\bibnamefont
  {Gimelshein}}, \bibinfo {author} {\bibfnamefont {L.}~\bibnamefont {Antiga}},
  \bibinfo {author} {\bibfnamefont {A.}~\bibnamefont {Desmaison}}, \bibinfo
  {author} {\bibfnamefont {A.}~\bibnamefont {Köpf}}, \bibinfo {author}
  {\bibfnamefont {E.}~\bibnamefont {Yang}}, \bibinfo {author} {\bibfnamefont
  {Z.}~\bibnamefont {DeVito}}, \bibinfo {author} {\bibfnamefont
  {M.}~\bibnamefont {Raison}}, \bibinfo {author} {\bibfnamefont
  {A.}~\bibnamefont {Tejani}}, \bibinfo {author} {\bibfnamefont
  {S.}~\bibnamefont {Chilamkurthy}}, \bibinfo {author} {\bibfnamefont
  {B.}~\bibnamefont {Steiner}}, \bibinfo {author} {\bibfnamefont
  {L.}~\bibnamefont {Fang}}, \bibinfo {author} {\bibfnamefont {J.}~\bibnamefont
  {Bai}}, \ and\ \bibinfo {author} {\bibfnamefont {S.}~\bibnamefont
  {Chintala}},\ }\href {\doibase 10.48550/arXiv.1912.01703} {\enquote {\bibinfo
  {title} {Pytorch: An imperative style, high-performance deep learning
  library},}\ } (\bibinfo {year} {2019}),\ \Eprint
  {http://arxiv.org/abs/1912.01703} {arXiv:1912.01703 [cs.LG]} \BibitemShut
  {NoStop}%
\bibitem [{\citenamefont {Schubert}\ \emph {et~al.}(2019)\citenamefont
  {Schubert}, \citenamefont {Neubert}, \citenamefont {P{\"o}schmann},\ and\
  \citenamefont {Protzel}}]{schubert2019circular}%
  \BibitemOpen
  \bibfield  {author} {\bibinfo {author} {\bibfnamefont {S.}~\bibnamefont
  {Schubert}}, \bibinfo {author} {\bibfnamefont {P.}~\bibnamefont {Neubert}},
  \bibinfo {author} {\bibfnamefont {J.}~\bibnamefont {P{\"o}schmann}}, \ and\
  \bibinfo {author} {\bibfnamefont {P.}~\bibnamefont {Protzel}},\ }in\
  \href@noop {} {\emph {\bibinfo {booktitle} {2019 IEEE intelligent vehicles
  symposium (IV)}}}\ (\bibinfo {organization} {IEEE},\ \bibinfo {year} {2019})\
  pp.\ \bibinfo {pages} {653--660}\BibitemShut {NoStop}%
\bibitem [{\citenamefont {Long}\ \emph {et~al.}(2015)\citenamefont {Long},
  \citenamefont {Shelhamer},\ and\ \citenamefont {Darrell}}]{long2015fully}%
  \BibitemOpen
  \bibfield  {author} {\bibinfo {author} {\bibfnamefont {J.}~\bibnamefont
  {Long}}, \bibinfo {author} {\bibfnamefont {E.}~\bibnamefont {Shelhamer}}, \
  and\ \bibinfo {author} {\bibfnamefont {T.}~\bibnamefont {Darrell}},\ }in\
  \href@noop {} {\emph {\bibinfo {booktitle} {Proceedings of the IEEE
  conference on computer vision and pattern recognition}}}\ (\bibinfo {year}
  {2015})\ pp.\ \bibinfo {pages} {3431--3440}\BibitemShut {NoStop}%
\bibitem [{\citenamefont {Cohen}\ and\ \citenamefont
  {Welling}(2016)}]{Cohen2016groupEquivariant}%
  \BibitemOpen
  \bibfield  {author} {\bibinfo {author} {\bibfnamefont {T.}~\bibnamefont
  {Cohen}}\ and\ \bibinfo {author} {\bibfnamefont {M.}~\bibnamefont
  {Welling}},\ }in\ \href {https://proceedings.mlr.press/v48/cohenc16.html}
  {\emph {\bibinfo {booktitle} {Proceedings of The 33rd International
  Conference on Machine Learning}}},\ \bibinfo {series} {Proceedings of Machine
  Learning Research}, Vol.~\bibinfo {volume} {48},\ \bibinfo {editor} {edited
  by\ \bibinfo {editor} {\bibfnamefont {M.~F.}\ \bibnamefont {Balcan}}\ and\
  \bibinfo {editor} {\bibfnamefont {K.~Q.}\ \bibnamefont {Weinberger}}}\
  (\bibinfo  {publisher} {PMLR},\ \bibinfo {address} {New York, New York,
  USA},\ \bibinfo {year} {2016})\ pp.\ \bibinfo {pages}
  {2990--2999}\BibitemShut {NoStop}%
\bibitem [{\citenamefont {Chung}\ \emph {et~al.}(2014)\citenamefont {Chung},
  \citenamefont {Gulcehre}, \citenamefont {Cho},\ and\ \citenamefont
  {Bengio}}]{Chung20141arXiv}%
  \BibitemOpen
  \bibfield  {author} {\bibinfo {author} {\bibfnamefont {J.}~\bibnamefont
  {Chung}}, \bibinfo {author} {\bibfnamefont {C.}~\bibnamefont {Gulcehre}},
  \bibinfo {author} {\bibfnamefont {K.}~\bibnamefont {Cho}}, \ and\ \bibinfo
  {author} {\bibfnamefont {Y.}~\bibnamefont {Bengio}},\ }\href
  {http://arxiv.org/abs/1412.3555} {\enquote {\bibinfo {title} {{Empirical
  Evaluation of Gated Recurrent Neural Networks on Sequence Modeling}},}\ }
  (\bibinfo {year} {2014}),\ \Eprint {http://arxiv.org/abs/1412.3555}
  {arXiv:1412.3555} \BibitemShut {NoStop}%
\bibitem [{\citenamefont {Shi}\ \emph {et~al.}(2015)\citenamefont {Shi},
  \citenamefont {Chen}, \citenamefont {Wang}, \citenamefont {Yeung},
  \citenamefont {Wong},\ and\ \citenamefont {Woo}}]{Shi2015NIPS}%
  \BibitemOpen
  \bibfield  {author} {\bibinfo {author} {\bibfnamefont {X.}~\bibnamefont
  {Shi}}, \bibinfo {author} {\bibfnamefont {Z.}~\bibnamefont {Chen}}, \bibinfo
  {author} {\bibfnamefont {H.}~\bibnamefont {Wang}}, \bibinfo {author}
  {\bibfnamefont {D.~Y.}\ \bibnamefont {Yeung}}, \bibinfo {author}
  {\bibfnamefont {W.~K.}\ \bibnamefont {Wong}}, \ and\ \bibinfo {author}
  {\bibfnamefont {W.~C.}\ \bibnamefont {Woo}},\ }\href@noop {} {\bibfield
  {journal} {\bibinfo  {journal} {Advances in Neural Information Processing
  Systems}\ }\textbf {\bibinfo {volume} {2015-Janua}},\ \bibinfo {pages} {802}
  (\bibinfo {year} {2015})},\ \Eprint {http://arxiv.org/abs/1506.04214}
  {arXiv:1506.04214} \BibitemShut {NoStop}%
\bibitem [{\citenamefont {Kingma}\ and\ \citenamefont
  {Ba}(2017)}]{kingmaadam2014}%
  \BibitemOpen
  \bibfield  {author} {\bibinfo {author} {\bibfnamefont {D.~P.}\ \bibnamefont
  {Kingma}}\ and\ \bibinfo {author} {\bibfnamefont {J.}~\bibnamefont {Ba}},\
  }\href {\doibase 10.48550/arXiv.1412.6980} {\enquote {\bibinfo {title} {Adam:
  A method for stochastic optimization},}\ } (\bibinfo {year} {2017}),\ \Eprint
  {http://arxiv.org/abs/1412.6980} {arXiv:1412.6980 [cs.LG]} \BibitemShut
  {NoStop}%
\bibitem [{\citenamefont {Bengio}\ \emph {et~al.}(2009)\citenamefont {Bengio},
  \citenamefont {Louradour}, \citenamefont {Collobert},\ and\ \citenamefont
  {Weston}}]{bengio2009curriculum}%
  \BibitemOpen
  \bibfield  {author} {\bibinfo {author} {\bibfnamefont {Y.}~\bibnamefont
  {Bengio}}, \bibinfo {author} {\bibfnamefont {J.}~\bibnamefont {Louradour}},
  \bibinfo {author} {\bibfnamefont {R.}~\bibnamefont {Collobert}}, \ and\
  \bibinfo {author} {\bibfnamefont {J.}~\bibnamefont {Weston}},\ }in\
  \href@noop {} {\emph {\bibinfo {booktitle} {Proceedings of the 26th annual
  international conference on machine learning}}}\ (\bibinfo {year} {2009})\
  pp.\ \bibinfo {pages} {41--48}\BibitemShut {NoStop}%
\bibitem [{Sta(2023)}]{StanfordBunny}%
  \BibitemOpen
  \href@noop {} {\enquote {\bibinfo {title} {Stanford university computer
  graphics laboratory, stanford bunny},}\ }\bibinfo {howpublished}
  {\url{https://graphics.stanford.edu/data/3Dscanrep/}} (\bibinfo {year}
  {2023})\BibitemShut {NoStop}%
\bibitem [{\citenamefont {Chen}\ and\ \citenamefont
  {Shen}(1998)}]{CHEN1998PFpseudospectral}%
  \BibitemOpen
  \bibfield  {author} {\bibinfo {author} {\bibfnamefont {L.}~\bibnamefont
  {Chen}}\ and\ \bibinfo {author} {\bibfnamefont {J.}~\bibnamefont {Shen}},\
  }\href {\doibase https://doi.org/10.1016/S0010-4655(97)00115-X} {\bibfield
  {journal} {\bibinfo  {journal} {Computer Physics Communications}\ }\textbf
  {\bibinfo {volume} {108}},\ \bibinfo {pages} {147} (\bibinfo {year}
  {1998})}\BibitemShut {NoStop}%
\bibitem [{\citenamefont {Kohn}\ and\ \citenamefont
  {Otto}(2002)}]{kohn2002upper}%
  \BibitemOpen
  \bibfield  {author} {\bibinfo {author} {\bibfnamefont {R.~V.}\ \bibnamefont
  {Kohn}}\ and\ \bibinfo {author} {\bibfnamefont {F.}~\bibnamefont {Otto}},\
  }\href@noop {} {\bibfield  {journal} {\bibinfo  {journal} {Communications in
  mathematical physics}\ }\textbf {\bibinfo {volume} {229}},\ \bibinfo {pages}
  {375} (\bibinfo {year} {2002})}\BibitemShut {NoStop}%
\bibitem [{\citenamefont {Kwon}\ \emph {et~al.}(2010)\citenamefont {Kwon},
  \citenamefont {Thornton},\ and\ \citenamefont
  {Voorhees}}]{kwon2010morphology}%
  \BibitemOpen
  \bibfield  {author} {\bibinfo {author} {\bibfnamefont {Y.}~\bibnamefont
  {Kwon}}, \bibinfo {author} {\bibfnamefont {K.}~\bibnamefont {Thornton}}, \
  and\ \bibinfo {author} {\bibfnamefont {P.}~\bibnamefont {Voorhees}},\
  }\href@noop {} {\bibfield  {journal} {\bibinfo  {journal} {Philosophical
  Magazine}\ }\textbf {\bibinfo {volume} {90}},\ \bibinfo {pages} {317}
  (\bibinfo {year} {2010})}\BibitemShut {NoStop}%
\end{thebibliography}%

\end{document}